\documentclass[12pt,english]{article}
\usepackage[T1]{fontenc}
\usepackage[latin9]{inputenc}
\usepackage{geometry}
\usepackage{color}
\usepackage{array}
\usepackage{refstyle}
\usepackage{float}
\usepackage{units}
\usepackage{amsmath}
\usepackage{amsthm}
\usepackage{amssymb}
\usepackage{graphicx}
\usepackage[numbers]{natbib}

\makeatletter

\AtBeginDocument{\providecommand\secref[1]{\ref{sec:#1}}}
\AtBeginDocument{\providecommand\subsecref[1]{\ref{subsec:#1}}}
\AtBeginDocument{\providecommand\figref[1]{\ref{fig:#1}}}
\AtBeginDocument{\providecommand\eqref[1]{\ref{eq:#1}}}
\providecommand{\tabularnewline}{\\}
\RS@ifundefined{subsecref}
  {\newref{subsec}{name = \RSsectxt}}
  {}
\RS@ifundefined{thmref}
  {\def\RSthmtxt{theorem~}\newref{thm}{name = \RSthmtxt}}
  {}
\RS@ifundefined{lemref}
  {\def\RSlemtxt{lemma~}\newref{lem}{name = \RSlemtxt}}
  {}

\numberwithin{equation}{section}
\numberwithin{figure}{section}

\usepackage[caption=false]{subfig}
\usepackage{url}
\usepackage{txfonts}
\let\omega\omegaup

\usepackage{amsfonts}
\usepackage{graphicx}
\usepackage{epstopdf}
\ifpdf
  \DeclareGraphicsExtensions{.eps,.pdf,.png,.jpg}
\else
  \DeclareGraphicsExtensions{.eps}
\fi

\@ifundefined{showcaptionsetup}{}{%
 \PassOptionsToPackage{caption=false}{subfig}}
\usepackage{subfig}
\makeatother

\usepackage{babel}
\begin{document}

\title{Synchronization of electrically coupled resonate-and-fire neurons}

\author{Thomas Chartrand\thanks{Graduate Group in Applied Mathematics and Center for Neuroscience,
University of California-Davis, Davis, CA (tmchartrand@ucdavis.edu).}\and Mark S. Goldman\thanks{Center for Neuroscience, Department of Neurobiology, Physiology and
Behavior, and Department of Ophthalmology and Vision Science, University
of California-Davis, Davis, CA}\and Timothy J. Lewis\thanks{Department of Mathematics, University of California-Davis, Davis,
CA}}
\maketitle
\begin{abstract}
Electrical coupling between neurons is broadly present across brain
areas and is typically assumed to synchronize network activity. However,
intrinsic properties of the coupled cells can complicate this simple
picture. Many cell types with strong electrical coupling have been
shown to exhibit resonant properties, and the subthreshold fluctuations
arising from resonance are transmitted through electrical synapses
in addition to action potentials. Using the theory of weakly coupled
oscillators, we explore the effect of both subthreshold and spike-mediated
coupling on synchrony in small networks of electrically coupled resonate-and-fire
neurons, a hybrid neuron model with linear subthreshold dynamics and
discrete post-spike reset. We calculate the phase response curve using
an extension of the adjoint method that accounts for the discontinuity
in the dynamics. We find that both spikes and resonant subthreshold
fluctuations can jointly promote synchronization. The subthreshold
contribution is strongest when the voltage exhibits a significant
post-spike elevation in voltage, or plateau. Additionally, we show
that the geometry of trajectories approaching the spiking threshold
causes a \textquotedbl{}reset-induced shear\textquotedbl{} effect
that can oppose synchrony in the presence of network asymmetry, despite
having no effect on the phase-locking of symmetrically coupled pairs.
\end{abstract}

\section{Introduction}

Synchronization of activity between neurons has been hypothesized
to contribute to a variety of brain functions \citep{singer1999NeuronalSynchronyVersatile},
including motor control \citep{dezeeuw1998Microcircuitryfunctioninferior},
memory \citep{fell2011rolephasesynchronization}, and coordination
between brain regions \citep{buzsaki2004NeuronalOscillationsCortical}.
This synchrony can be supported by either electrical or chemical synapses,
or some combination of the two. Because electrical synapses (gap junctions)
diffusively couple the voltages of connected cells, their effect is
typically thought to be synchronizing, an idea with support from both
theoretical and experimental studies \citep{bennett2004ElectricalCouplingNeuronal,connors2004ELECTRICALSYNAPSESMAMMALIAN,mancilla2007SynchronizationElectricallyCoupled}.
However, their effect is potentially more complex, in part because
the coupling combines effects at the dramatically different timescales
of spiking activity and subthreshold fluctuations of membrane potential
\citep{connors2004ELECTRICALSYNAPSESMAMMALIAN,lewis2003Dynamicsspikingneurons,chow2000DynamicsSpikingNeurons,pfeuty2003ElectricalSynapsesSynchrony}.
Observations have shown that many cell types with strong electrical
coupling exhibit resonant properties, which can create distinctive
voltage fluctuations between spikes \citep{hutcheon2000Resonanceoscillationintrinsic}.
The same intrinsic properties that determine a frequency-selective
spiking response to input also cause subthreshold dynamics such as
transient oscillations, hyperpolarization followed by rebound, or
a depolarized plateau following the spike; any of these effects can
potentially contribute to synchronization. Our goal is to explore
the interaction between electrical coupling and resonant intrinsic
dynamics of spiking neurons, to understand both the dynamical mechanisms
involved and their relevance to the function of neural systems.

We study the effect of electrical coupling on the synchronization
of resonant spiking neurons, by applying the theory of weakly coupled
oscillators to reduce the complexity of the synchronization problem
and gain analytical insight \citep{ashwin2016MathematicalFrameworksOscillatory}.
This technique relies on a perturbative approximation to derive a
reduced \emph{phase model} for limit cycle oscillators \citep{schwemmer2012TheoryWeaklyCoupled}.
Synchronization of the phase model is determined by the \emph{interaction
function}, which captures the effect of coupling as a function of
the phase of each oscillator along its periodic limit cycle. Determining
how the interaction function depends on a property of the oscillator,
such as subthreshold resonance, spike size, or post-spike behavior,
shows how that property contributes to synchronization. Common challenges
in phase reduction analysis are that it may not be possible to independently
vary the dynamical properties of interest, or to analytically compute
the interaction function.

For the phase reduction to be analytically tractable, we use a minimal
hybrid neuron model for the dynamics of resonant spiking. Hybrid neuron
models, such as the integrate-and-fire model and its generalizations,
idealize spiking as a threshold crossing with discrete post-spike
reset, combined with continuous subthreshold dynamics between spikes.
We focus on the resonate-and-fire model \citep{izhikevich2001Resonateandfireneurons},
which has linear damped oscillations as its subthreshold dynamics.
The simplification of spiking allows the model to remain analytically
tractable, while the discrete reset map helps to create complex resonant
or integrator-like spiking dynamics. With the separation of discrete
and continuous dynamics, we can independently vary the subthreshold
and spiking properties of the model and determine their effects on
the synchrony of small model networks. 

On the other hand, the discrete reset map complicates the application
of weakly coupled oscillator theory for the analysis of synchrony.
The reset creates a discontinuity of the limit cycle, breaking a standard
assumption in one step of the phase model reduction: calculation of
the phase response curve. The \emph{phase response curve} (PRC) measures
the phase shift resulting from a perturbation to the oscillator at
any point along the limit cycle. The discontinuous spiking dynamics
lead to discontinuity of the PRC. In certain cases, discontinuous
PRCs can be calculated directly \citep{miura2006Globallycoupledresonateandfire,coombes2009GapJunctionsEmergent,ermentrout2012PhaseResponseCurves},
but the discontinuity in general necessitates an extension of standard
methods for calculating the PRC. Shirasaka et al. \citep{shirasaka2017Phasereductiontheory}
recently proved that for general hybrid models the PRC can be calculated
by a variation to the standard ``adjoint method,'' as previously
suggested by Ladenbauer et al. \citep{ladenbauer2012ImpactAdaptationCurrents}.
We present an alternative, intuitive derivation of this result, elucidating
the connection to the geometry of the threshold and reset in the context
of hybrid neuron models, and apply this understanding to our resonate-and-fire
analysis.

The paper is organized as follows. In \secref{Resonate-and-fire-model},
we describe the general properties and history of hybrid models in
neuroscience, and define our generalization of the resonate-and-fire
model. In \secref{Phase-reduce}, we review the theory of weakly coupled
oscillators and present our approach to calculating the PRC for hybrid
models. The remainder of the paper contains our analysis of synchronization
in the resonate-and-fire model. In \secref{Resonate-and-fire-phase-reductio},
we apply our adjoint method approach to obtain an analytical expression
for the PRC and interaction function of the electrically coupled resonate-and-fire
neuron. To explore the dependence of the interaction function on model
parameters, we focus separately on the even- and odd-symmetric components
of the interaction function, which lead to distinct effects on synchrony.
In \secref{odd-pair}, we show that the spike itself always promotes
robust synchrony through its contributions to the odd component, while
the subthreshold fluctuations can additionally strongly promote synchrony
in a ``plateau potential'' regime with strong resonant dynamics.
We also show, in \secref{Even-3}, that the threshold and reset can
phase shift the interaction function through \textquotedbl{}reset-induced
shear\textquotedbl{} arising from the geometry of trajectories crossing
the threshold, leading to an even component that can have complex
effects on synchronization. In example three-cell networks, the presence
of this even component leads to often-ignored effects on network synchronization
when heterogeneity of frequencies or coupling breaks the symmetry
of the interactions. 

\section{Resonate-and-fire model\label{sec:Resonate-and-fire-model}}

\subsection{Hybrid models\label{subsec:Hybrid-models}}

We first introduce some fundamental examples of hybrid models in neuroscience,
and provide notation that will be used in the following sections.
Hybrid models have a central role in the history of mathematical neuroscience.
Well before the detailed processes generating action potentials were
understood, Lapicque \citep{lapicque1907Recherchesquantitativesexcitation}
postulated that inputs to a neuron accumulate in a continuous process
of integration, eventually triggering a spike. This idea led to the
leaky integrate-and-fire model (\ref{eq:LIF}) and a number of variations
that are still widely used \citep{ermentrout1986ParabolicBurstingExcitable,izhikevich2001Resonateandfireneurons,izhikevich2003Simplemodelspiking,brette2005AdaptiveExponentialIntegrateandFire,abbott1999Lapicqueintroductionintegrateandfire}.
The separation of the dramatically different timescales of spiking
and subthreshold dynamics into distinct mechanisms gives these models
both computational efficiency and analytic tractability. Generalized
integrate-and-fire models are remarkably effective at reproducing
diverse spiking behaviors \citep{izhikevich2003Simplemodelspiking,izhikevich2004Whichmodeluse,mihalas2009GeneralizedLinearIntegrateandFire}
and can even be fit directly to spike trains \citep{jolivet2004GeneralizedIntegrateandFireModels}. 

Hybrid models have surprisingly rich and complex dynamics, inspiring
active study from both neuroscience and dynamical systems perspectives
\citep{touboul2009SpikingDynamicsBidimensional,schwemmer2012BistabilityLeakyIntegrateandFire,carmona2013Melnikovtheoryclass}.
For the analysis of synchrony, a number of studies have considered
networks of coupled integrate-and-fire neurons and single-variable
variants \citep{ermentrout1981Phaselockingweaklycoupled,mirollo1990SynchronizationPulseCoupledBiological,kuramoto1991Collectivesynchronizationpulsecoupled,coombes1999ModelockingArnold,golomb2000NumberSynapticInputs,neltner2000SynchronyHeterogeneousNetworks,lewis2003Dynamicsspikingneurons,pfeuty2003ElectricalSynapsesSynchrony,ostojic2009Synchronizationpropertiesnetworks,lewis2012UnderstandingActivityElectrically,luccioli2012CollectiveDynamicsSparse}.
However, despite the importance of synchronization in neural dynamics,
only a few studies have addressed the synchronization of more complex
hybrid models with more than a single variable, which is necessary
to exhibit resonance \citep{digarbo2001Dynamicalbehaviorlinearized,miura2004Pulsecoupledresonateandfiremodels,miura2006Globallycoupledresonateandfire,coombes2012Nonsmoothdynamicsspiking,ermentrout2012PhaseResponseCurves,ladenbauer2012ImpactAdaptationCurrents,ladenbauer2013Adaptationcontrolssynchrony}.

The leaky integrate-and-fire model consists of a single voltage-like
variable with linear subthreshold dynamics between spikes \citep{lapicque1907Recherchesquantitativesexcitation,brunel2007Quantitativeinvestigationselectrical,abbott1999Lapicqueintroductionintegrateandfire}.
External current input $I\left(t\right)$ is integrated through changes
in the neuron's membrane potential (voltage) subject to a ``leak,''
or linear decay over time. When voltage crosses a threshold $v_{T}$
from below, a spike occurs and the voltage is reset to $v_{R}$.
\begin{gather}
\frac{dv}{dt}=-v+I\left(t\right),\ v\left(t^{-}\right)=v_{T}\implies v\left(t^{+}\right)=v_{R}.\label{eq:LIF}
\end{gather}
For sufficiently large constant current input, the equilibrium voltage
is pushed over the threshold, $I\left(t\right)=v_{eq}>v_{T}$, and
the model exhibits a regular spiking state, a limit cycle with periodic
firing. Note that the existence of a limit cycle in one dimension
is enabled by the discontinuous reset.

A number of variations on this model have been proposed. Some models
add nonlinear subthreshold dynamics in the form of a quadratic \citep{ermentrout1986ParabolicBurstingExcitable,brunel2003Firingratenoisy},
exponential \citep{fourcaud-trocme2003HowSpikeGeneration}, or other
function to more accurately model the approach to threshold. Adding
a second ``adaptation'' variable \citep{treves1993Meanfieldanalysisneuronal,richardson2003SubthresholdFiringRateResonance}
allows the model to reflect various slow recovery processes in the
neuron (typically the gating dynamics of ion channels) that can create
resonance and adaptation. With the presence of a second variable,
not only can the subthreshold dynamics be more complex, but the threshold
voltage $v_{T}$ becomes a \emph{threshold manifold}, and the reset
voltage $v_{R}$ can generalize to a \emph{reset map} from this threshold
manifold to a corresponding \emph{reset manifold}. While models with
a ``hard reset,'' like (\ref{eq:LIF}), map all trajectories to
a single point, assuming the adaptation process saturates during the
spike and eliminates any history in the dynamics, models with a ``soft
reset,'' like the adaptive exponential integrate-and-fire model \citep{brette2005AdaptiveExponentialIntegrateandFire},
instead map to a line of constant voltage. Generally, the soft reset
increments the adaptation variable, modeling a rapid change in the
state of ion channels which, unlike the hard reset, allows changes
in the adaptation variable to persist over multiple cycles.

Although a hybrid neuron model can in general consist of any number
of dimensions with complex threshold manifolds and reset maps \citep{mihalas2009GeneralizedLinearIntegrateandFire},
a two-dimensional model is sufficient to illustrate the essential
properties. For our explanations, we restrict models to two dimensions
$x=\left(v,\,w\right)$ with threshold manifold $\mathcal{T}$, reset
manifold $\mathcal{R}$, and corresponding reset map $R:\,\mathcal{T}\mapsto\mathcal{R}$.
\begin{gather}
\frac{dx}{dt}=\frac{d}{dt}\begin{pmatrix}v\\
w
\end{pmatrix}=f\left(x\right),\ x\left(t^{-}\right)\in\mathcal{T}\implies x\left(t^{+}\right)=R\left(x\left(t^{-}\right)\right).\label{eq:gen_hybrid}
\end{gather}

\subsection{Resonate-and-fire model}

The basic elements of the resonate-and-fire model were developed independently
by some of the earliest mathematical neuroscientists, Rashevsky \citep{rashevsky1933Outlinephysicomathematicaltheory}
and Hill \citep{hill1936ExcitationAccommodationNerve}. Izhikevich
\citep{izhikevich2001Resonateandfireneurons} introduced a modern
version of the model and coined the ``resonate-and-fire'' name.
The model's response dynamics have been shown to reproduce essential
features of the dynamics of resonant cells, including a frequency-selective
firing response to periodic input and a rebound of voltage following
hyperpolarizing input \citep{izhikevich2001Resonateandfireneurons}.
These properties have been widely observed in many classes of neurons,
and can be caused by a variety of voltage-dependent ion channels \citep{hutcheon2000Resonanceoscillationintrinsic}.
Note that since rebound from hyperpolarization requires crossing of
subthreshold voltage trajectories, our model must have more then one
dimension. The same holds for non-monotonic voltage trajectories with
a post-spike plateau potential. The linear two-dimensional resonate-and-fire
model is thus a minimal model for studying these phenomena. The resonate-and-fire
model has previously been studied with periodic forcing \citep{khajehalijani2009Modelockingperiodically}
and in the context of pulse-coupled pairs of neurons \citep{miura2004Pulsecoupledresonateandfiremodels}
and globally pulse-coupled populations \citep{miura2006Globallycoupledresonateandfire}. 

The subthreshold dynamics of the resonate-and-fire model are linear
in two dimensions, consisting of linear decay of both the voltage
variable and the adaptation variable towards an equilibrium voltage
$v_{eq}$, combined with coupling between the variables to create
linear damped oscillations.

\begin{equation}
\frac{dx}{dt}=\frac{d}{dt}\begin{pmatrix}v\\
w
\end{pmatrix}=\omega\begin{pmatrix}-\lambda & -1\\
1 & -\lambda
\end{pmatrix}\begin{pmatrix}v-v_{eq}\\
w
\end{pmatrix}.\label{eq:resfire_linear}
\end{equation}
The resonate-and-fire hard reset rule specifies that if a threshold
$v_{T}$ in the voltage coordinate is crossed from below, the state
is instantaneously reset to a single point $x_{R}$. 
\begin{equation}
v\left(t^{-}\right)=v_{T}\implies x\left(t^{+}\right)=R\left(x\left(t^{-}\right)\right)=x_{R}=\begin{pmatrix}v_{R}\\
w_{R}
\end{pmatrix}.\label{eq:resfire_reset}
\end{equation}

\begin{figure}
\begin{centering}
\includegraphics[scale=0.7]{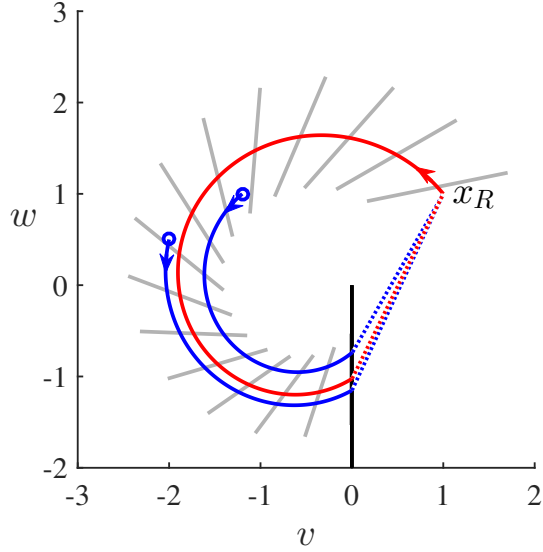}
\par\end{centering}
\caption{\label{fig:rf_pp}Spiking limit cycle of the resonate-and-fire model
with hard reset, shown in the $\left(v,\,w\right)$ phase plane. Two
nearby trajectories (blue) converge to the limit cycle (red) immediately
following the instantaneous reset (dotted lines) at the reset point
$x_{R}$. Isochrons (gray) connect points of the same asymptotic phase
(including along the spiking threshold at $v_{T}=0$). Parameters:
$\lambda=0.1$ $w_{R}=1$, $v_{R}=1$, $v_{eq}=-0.5$.}
\end{figure}

\begin{figure}
\begin{centering}
\includegraphics{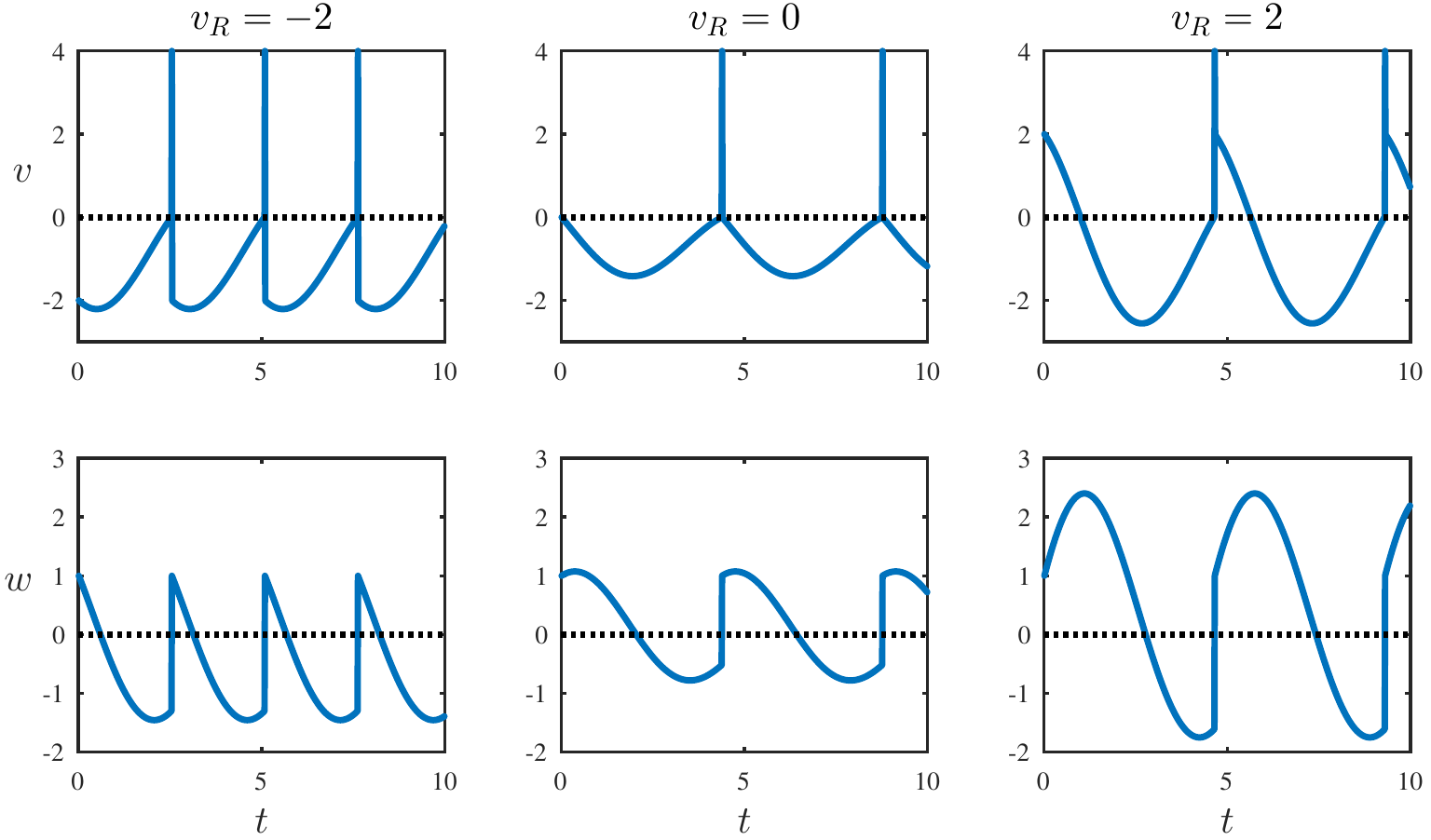}
\par\end{centering}
\caption{\label{fig:rf_traj}Trajectories of the $v-$ and $w-$component limit
cycles for varying values of the reset voltage $v_{R}$, showing examples
of ``after-hyperpolarization'' (AHP) with strongly negative reset
voltage, (left), ``plateau potential'' with elevated reset voltage
(right), and an intermediate trajectory (center). Instantaneous spikes
are added on to the limit cycle (\ref{eq:rf_lc_both}) at each threshold
crossing. Parameters: $\lambda=0.1$ $w_{R}=1$, $v_{eq}=-0.5$.}
\end{figure}

We set the threshold voltage to an arbitrary value $v_{T}=0$, rescale
time such that $\omega=1$, and unless otherwise specified choose
a fixed value for the decay parameter, $\lambda=0.1$. Our results
are qualitatively similar as long as $\lambda$ is sufficiently small.
In the opposite extreme, taking the limit $\lambda\to\infty$ with
the product $\lambda\omega$ fixed recovers the leaky integrate-and-fire
model.\footnote{We also note that taking this limit in our phase-reduced model as
expressed in \subsecref{app-Subthreshold-interaction-functio} recovers
known results for the phase reduction of the leaky integrate-and-fire
model \citep{lewis2012UnderstandingActivityElectrically}.} With these parameters fixed, the reset parameters and equilibrium
voltage determine the existence and properties of spiking in the model.
Varying the equilibrium voltage $v_{eq}$ relative to the threshold
reflects a combined effect of altering the biophysical resting potential
and the tonic (constant) current input to the cell. If a trajectory
starting from the reset point $x_{R}$ crosses the spiking threshold
and is again reset, it forms a spiking limit cycle corresponding to
regular spiking, shown in \figref{rf_pp}. We explore more detailed
existence conditions in \subsecref{spiking-limit-cycles}, but a limit
cycle will always exist for sufficiently large constant current input,
moving the equilibrium voltage $v_{eq}$ over the spiking threshold
just as in the integrate-and-fire model. The reset parameters, especially
the reset voltage $v_{R}$, also have a strong effect on the shape
of the voltage trajectory for this limit cycle, as shown in \figref{rf_traj}.
The voltage trace for an oscillator reset near the peak of its oscillation
(large positive reset $v_{R}$, \figref{rf_traj} right) resembles
a plateau potential, a sustained post-spike elevation in voltage.
The trace for an oscillator reset near its trough (large negative
reset $v_{R}$, \figref{rf_traj} left) resembles a fast after-hyperpolarization
(AHP), a post-spike dip in voltage. We will refer to these strong
negative and positive reset regimes simply as ``plateau'' and ``AHP''
below. An intermediate case with $v_{R}=0$ is shown in \figref{rf_traj}
center. As several studies have pointed out, these distinct after-potential
regimes can significantly affect synchronization mediated by electrical
synapses \citep{chow2000DynamicsSpikingNeurons,pfeuty2003ElectricalSynapsesSynchrony,mancilla2007SynchronizationElectricallyCoupled},
which we will investigate further using the resonate-and-fire model. 

The regular spiking solution for the resonate-and-fire model is defined
to start from the reset point at time $t=0$ and is valid for times
up to the period $T$, when the trajectory crosses threshold.
\begin{equation}
\begin{aligned}\bar{v}\left(t\right) & =v_{eq}+r_{0}e^{-\lambda t}\cos\left(t+\theta_{0}\right),\\
\bar{w}\left(t\right) & =r_{0}e^{-\lambda t}\sin\left(t+\theta_{0}\right),
\end{aligned}
\label{eq:rf_lc_both}
\end{equation}
where $\left(r_{0},\,\theta_{0}\right)$ are the polar coordinates
of the reset point relative to equilibrium, such that $\left(v_{R},\,w_{R}\right)=\left(v_{eq}+r_{0}\cos\theta_{0},\,r_{0}\sin\theta_{0}\right)$.

\subsection{Extended resonate-and-fire model: soft reset, spikes, and coupling}

From the perspective of a general hybrid model, the spike-and-reset
dynamics restrict the definition of both the threshold and reset:
with the hard reset map $R\left(w\right)=x_{R}$ from (\ref{eq:resfire_reset}),
the threshold manifold is a line of constant voltage $\mathcal{T}=\left\{ \left(v,\,w\right):\,v=v_{T}\right\} $,
and the reset manifold is a single point $\mathcal{R}=\left\{ x_{R}=\left(v_{R},\,w_{R}\right)\right\} $.
In order to expand the possible dynamics of resonate-and-fire models
while retaining simplicity, we also consider the ``soft reset''
variation commonly used in other hybrid models, which increments the
adaptation variable by a constant value. The reset manifold for the
soft reset is thus a line of constant voltage $\mathcal{R}=\left\{ \left(v,\,w\right):\,v=v_{R}\right\} $.

\begin{equation}
x\left(t^{+}\right)=R\left(w\left(t^{-}\right)\right)=\left(v_{R},\,w\left(t^{-}\right)+\Delta w\right).\label{eq:softreset}
\end{equation}
More generally, the threshold and reset manifolds could have arbitrary
orientations and locations - we briefly discuss considerations for
the general case in \subsecref{RF-PRC}. 

To build a network model of resonate-and-fire oscillators, we define
the electrical coupling as direct exchange of current, coupling the
voltage but not the adaptation variables \citep{lewis2012UnderstandingActivityElectrically}.
Specifically, we assume the coupling to be linear in the voltage difference
and let $k_{ij}$ be the coupling strength between neuron $j$ and
neuron $i$. For the cells in the network, we introduce heterogeneity
of intrinsic frequencies $\omega_{i}$ about a mean frequency $\bar{\omega}=1$,
while keeping other parameters fixed for the population. The resulting
network model is as follows, where $I_{c}\left(x_{i},\,x_{j}\right)=v_{j}-v_{i}$.

\begin{equation}
\frac{d}{dt}\begin{pmatrix}v_{i}\\
w_{i}
\end{pmatrix}=\omega_{i}\begin{pmatrix}-\lambda & -1\\
1 & -\lambda
\end{pmatrix}\begin{pmatrix}v_{i}-v_{eq}\\
w_{i}
\end{pmatrix}+\begin{pmatrix}\sum_{j}k_{ij}I_{c}\left(x_{i},\,x_{j}\right)\\
0
\end{pmatrix}.\label{eq:resfire net}
\end{equation}

The spiking dynamics of the resonate-and-fire model (\ref{eq:resfire_linear}-\ref{eq:softreset})
describe the subthreshold dynamics, threshold crossing, and post-spike
reset but not the spike itself. In order to model the effects of electrical
coupling, which transfers current based on voltage fluctuations both
during and between spikes, we must supplement the model by a description
of the transient voltage spike. We choose to define the spike as a
$\delta$-function at the instant of crossing threshold \citep{lewis2012UnderstandingActivityElectrically},
\begin{equation}
\bar{v}\left(t\right)=v_{eq}+r_{0}e^{-\lambda t}\cos\left(t+\theta_{0}\right)+M\delta\left(t\right),\ t\in\left[0,\,T\right]\label{eq:resfire trajectory}
\end{equation}
The spike magnitude parameter $M$ determines the integral of the
spike over time, or current exchanged. When $M$ is large relative
to the subthreshold fluctuations, the effect of electrical coupling
is essentially pulse coupling, as studied by Miura and Okada \citep{miura2004Pulsecoupledresonateandfiremodels}.
(However, their results are not directly comparable, as they consider
pulse coupling to the adaptation ($w$) variable rather than the voltage
($v$) variable.) We focus on a regime where the spike effect is comparable
to or smaller than the subthreshold fluctuations, and use the theory
of weakly coupled oscillators to analyze the spike and subthreshold
contributions to synchrony independently.

\section{Phase reduction theory\label{sec:Phase-reduce}}

\subsection{Phase mapping for weakly coupled oscillators}

The theory of weakly coupled oscillators is a method of model reduction
with the goal of creating simpler models for the dynamics of interacting
limit-cycle oscillators. The state of each oscillator is mapped to
a single variable: the phase (or timing) of oscillations. This process
is referred to as ``phase reduction'' and the result as a ``phase
model.'' The oscillators' intrinsic dynamics and the coupling between
oscillators together determine the \emph{interaction function}, which
captures the modulation of instantaneous frequency caused by coupling.
This scalar function defines the coupling between the phase oscillators,
and thus determines the dynamics of the coupled system (together with
any heterogeneity of intrinsic frequencies). Dynamical properties
such as the stability of phase-locked states can be directly assessed
from features of the interaction function, as we will show in \secref{Even-3}.
Although strictly derived in the limit of weak coupling and heterogeneity,
predictions from the phase model often remain valid even at moderate
levels of coupling. Here we first provide a brief derivation of the
phase model for a continuous coupled oscillator system. We then explain
the specific challenges presented by hybrid systems and our approach
to overcoming these challenges.

The dynamics of a general system of coupled oscillators are described
by
\begin{equation}
\frac{dx_{i}}{dt}=f\left(x_{i}\right)+g_{i}\left(x_{i}\right)+\sum_{j}k_{ij}I_{c}\left(x_{i},\,x_{j}\right).\label{eq:pop_model_generic}
\end{equation}
The state space of each individual oscillator is $x_{i}\in\mathbb{R}^{n}$,
where $n$ is the dimension of the oscillator model. The intrinsic
dynamics of each oscillator are defined by $f+g_{i}$, where $f$
gives an average of the intrinsic dynamics across all oscillators,
and $g_{i}$ captures the heterogeneity of the oscillators. The average
dynamics must have an asymptotically stable $T$-periodic limit cycle
$\bar{x}\left(t\right)$ defined by 
\begin{equation}
\frac{d\bar{x}}{dt}=f\left(\bar{x}\right),\ \bar{x}(t+T)=\bar{x}\left(t\right).\label{eq:dynamics_mean}
\end{equation}
The pairwise coupling is defined by the coupling function $I_{c}$,
which is weighted by connection strengths $k_{ij}$ and summed over
all pairs to give the total coupling. The phase reduction requires
the assumption of weak coupling and weak heterogeneity, meaning that
the heterogeneity $g_{i}$ and the total coupling term must both be
small compared to the average intrinsic dynamics $f$.

To reduce the model for this coupled system, we define a \emph{phase
mapping} $\Theta:\ \mathbb{R}^{n}\mapsto\mathbb{R}$, from the state
$x$ to a scalar phase variable $\theta$ that uniquely identifies
points on the limit cycle. The dynamics of the coupled system are
translated through this mapping to define the phase model, 
\[
\frac{d\theta_{i}}{dt}=\Omega_{i}+\sum_{j}k_{ij}H\left(\theta_{j}-\theta_{i}\right).
\]
The dynamics of this reduced system depend only on the interaction
function $H$ and the heterogeneity of frequencies $\Omega_{i}$.
Below, we show how these are derived from the coupling $I_{c}$ and
heterogeneity $g_{i}$ of the full model.

We first define the phase map for points on the limit cycle, giving
a periodic ``time'' coordinate, 
\begin{equation}
\theta=\Theta\left(\bar{x}\left(t\right)\right)\equiv t.\label{eq:phase_def}
\end{equation}
Phase is unique modulo $T$, with $\theta=0$ determined by our choice
of initial condition $\bar{x}\left(0\right)$ for the reference limit
cycle. (Note that phase is sometimes rescaled to a period of 1 or
$2\pi$, but we choose to keep the natural units of time.)

The phase map can then be extended beyond the limit cycle to give
the ``asymptotic phase'' of all points in the basin of attraction.
Trajectories that eventually converge are assigned the same phase,
i.e.,
\begin{equation}
\Theta\left(x\left(t\right)\right)=\Theta\left(\bar{x}\left(\theta\right)\right)=\theta\text{ iff }\lim_{t'\to\infty}\left[x\left(t+t'\right)-\bar{x}\left(\theta+t'\right)\right]=0.\label{eq:asymp_phase}
\end{equation}
Although typically not calculated explicitly, the full phase mapping
can be visualized by its level surfaces, referred to as isochrons,
linking points that approach the limiting trajectory with the same
timing (\figref{rf_pp}).

\subsection{Phase response curve and the adjoint method\label{subsec:Phase-response-curve}}

The full phase mapping is difficult to find analytically in all but
the simplest contexts. Fortunately, the weak coupling assumption allows
the phase reduction to proceed with a linear approximation of the
mapping about the limit cycle $\bar{x}\left(t\right)$, which is much
easier to compute. For a trajectory close to the limit cycle, phase
can be approximated as linearly dependent on the deviation away from
the limit cycle, $\Delta x\left(t\right)$. 
\begin{equation}
\Theta\left(x\left(t\right)\right)=\Theta\left(\bar{x}\left(t\right)+\Delta x\left(t\right)\right)\approx\Theta\left(\bar{x}\left(t\right)\right)+\nabla\Theta\left(\bar{x}\left(t\right)\right)^{T}\Delta x\left(t\right).\label{eq:phase_lin}
\end{equation}
The \emph{infinitesimal} phase response curve $Z$ (iPRC or PRC, also
called the ``phase sensitivity function'') is defined as the proportional
shift in phase caused by infinitesimal perturbations to the limit
cycle,
\begin{equation}
Z\left(\theta\right)=\nabla\Theta\left(\bar{x}\left(\theta\right)\right).\label{eq:prc_def}
\end{equation}
Note that the PRC is naturally defined as a vector-valued function,
giving the effect of perturbations to each state variable. In the
context of neural synchrony, however, the voltage component is usually
most important, because perturbations tend to be currents and thus
directly affect only the current balance equation for the dynamics
of voltage.

A direct method to approximate the PRC either experimentally or computationally
is simply to measure phase shifts caused by many small but finite
perturbations. We instead follow the ``adjoint method,'' which derives
and solves a differential equation for the PRC tied to the limit-cycle
dynamics. Below we provide a brief exposition that captures the essence
of this method and its proof. 

From the definition of asymptotic phase (\ref{eq:asymp_phase}), the
phase difference between the limit cycle $\bar{x}\left(t\right)$
and a nearby trajectory $x\left(t\right)$ must be constant in time.
That is, the separation $\Delta x\left(t\right)=x\left(t\right)-\bar{x}\left(t\right)$
must satisfy
\begin{gather}
c=\Theta\left(\bar{x}\left(t\right)+\Delta x\left(t\right)\right)-\Theta\left(\bar{x}\left(t\right)\right)\approx Z\left(t\right)^{T}\Delta x(t)\nonumber \\
0\approx\frac{d}{dt}\left(Z\left(t\right)^{T}\Delta x(t)\right).\label{eq:adjoint_step1}
\end{gather}
To first order, the deviation from the limit cycle $\Delta x\left(t\right)$
evolves according to the Jacobian matrix of derivatives of the system
dynamics, $Df$, evaluated on the limit cycle given by (\ref{eq:dynamics_mean}).
\[
\frac{d}{dt}\Delta x\left(t\right)\approx Df\left(\bar{x}\left(t\right)\right)\Delta x\left(t\right).
\]
Substituting this expression into (\ref{eq:adjoint_step1}), we obtain
\[
\left(\frac{dZ(t)}{dt}^{T}+Z\left(t\right)^{T}Df\left(\bar{x}\left(t\right)\right)\right)\Delta x(t)=0.
\]
Because this holds for any $\Delta x$, $Z$ must satisfy the following
$T$-periodic linear differential equation known as the adjoint equation,
defined by the adjoint of the linearized limit-cycle dynamics,
\begin{equation}
\frac{dZ(t)}{dt}+Df\left(\bar{x}\left(t\right)\right)^{T}Z\left(t\right)=0.\label{eq:adjoint}
\end{equation}
The PRC is the unique periodic solution of (\ref{eq:adjoint}), given
a normalization constraint that follows from our definition of phase
on the limit cycle (\ref{eq:phase_def}).
\begin{equation}
1=\frac{d}{dt}\Theta\left(\bar{x}\left(t\right)\right)=Z\left(t\right)^{T}f\left(\bar{x}\left(t\right)\right).\label{eq:normalize}
\end{equation}
Note that if this constraint is satisfied at any single time, (\ref{eq:adjoint})
ensures it will remain satisfied for all time.

\subsection{The phase model and interaction function}

Using the PRC result derived via the adjoint method, we can complete
the derivation of the phase reduction of the coupled oscillator system
(\ref{eq:pop_model_generic}). The effect of any weak time-dependent
perturbation on the phase of an oscillator, including the effect of
coupling, is governed by the PRC. Specifically, if the dynamics of
the limit cycle are continuously (weakly) perturbed according to $\frac{dx}{dt}=f\left(x\right)+\epsilon p\left(t\right)$
(for $\epsilon\ll1$), the perturbed phase $\theta=\Theta\left(x\left(t\right)\right)$
satisfies 
\[
\frac{d\theta}{dt}=\nabla\Theta^{T}\frac{dx}{dt}\approx1+\epsilon Z\left(t\right)^{T}p\left(t\right).
\]
Since the perturbation is weak, its effects occur on a slow timescale,
$O\left(\epsilon\right)$, which can be separated from the faster
dynamics of unperturbed phase, $\frac{d\theta}{dt}=1$. If the perturbation
is also periodic with the intrinsic period $T$, this separation allows
us to eliminate the explicit time-dependence $p\left(t\right)$ by
averaging the slow effect of coupling over a full period of the fast
phase dynamics. 

Because the perturbations that define the coupled population model
(\ref{eq:pop_model_generic}) are close to periodic, their effects
can be approximated by the method of averaging.\footnote{A more detailed explanation of the phase reduction can be derived
from singular perturbation theory or averaging theory \citep{ermentrout1991Multiplepulseinteractions,schwemmer2012TheoryWeaklyCoupled}.} The heterogeneity and coupling perturbations to an oscillator are
functions of its own trajectory $x_{i}$ and those of the other oscillators
$x_{j}$. Each trajectory is approximated by the $T$-periodic average
limit cycle, $x_{i}\approx\bar{x}\left(\theta_{i}\right)$ (where
$\theta_{i}=\Theta\left(x_{i}\right)$); therefore we can approximate
these perturbations as periodic functions of phase, $g_{i}\left(x_{i}\right)\approx\tilde{g_{i}}\left(\theta_{i}\right)\equiv g_{i}\left(\bar{x}\left(\theta_{i}\right)\right)$
and likewise for $\tilde{I}_{c}$. 
\begin{eqnarray*}
\frac{dx_{i}}{dt} & = & f\left(x_{i}\right)+\overset{\epsilon p\left(t\right)}{\overbrace{\left(\tilde{g_{i}}\left(\theta_{i}\right)+\sum_{j}k_{ij}\tilde{I}_{c}\left(\theta_{i},\,\theta_{j}\right)\right)}}
\end{eqnarray*}
The final result of the phase reduction is the phase model, i.e.
\begin{equation}
\frac{d\theta_{i}}{dt}=\Omega_{i}+\sum_{j}k_{ij}H\left(\theta_{j}-\theta_{i}\right),\label{eq:phase_model}
\end{equation}
where
\begin{equation}
\Omega_{i}=1+\frac{1}{T}\int_{0}^{T}Z\left(t\right)^{T}\tilde{g_{i}}\left(t\right)dt,\label{eq:phase_het}
\end{equation}
\begin{equation}
H\left(\theta_{j}-\theta_{i}\right)=\frac{1}{T}\int_{0}^{T}Z\left(t\right)^{T}\tilde{I}_{c}\left(t,\,t+\theta_{j}-\theta_{i}\right)dt.\label{eq:phase_interaction}
\end{equation}
The frequency term $\Omega_{i}$ arises from the intrinsic heterogeneity
in cellular properties, and the interaction function $H$ from the
coupling. We note that the form of $H$ as a function of phase differences,
$\phi_{j-i}=\theta_{j}-\theta_{i}$, arises from the application of
averaging.

\subsection{PRC for hybrid models\label{subsec:Hybrid-PRC}}

For hybrid models, the PRC as well as the trajectory may be discontinuous
at the threshold crossing. Without the periodicity constraint acting
as a boundary condition, the adjoint equation (\ref{eq:adjoint})
with normalization (\ref{eq:normalize}) no longer has a unique solution,
and a naive application of the adjoint method fails to find the PRC.
Intuitively, the resolution is to find the appropriate ``reset''
map or boundary condition for the PRC. Shirasaka et al. \citep{shirasaka2017Phasereductiontheory}
present a boundary condition linked to the \emph{saltation matrix},
a correction to the linearized dynamics of a hybrid system to account
for discontinuity across a boundary \citep{bernardo2008Piecewisesmoothdynamicalsystems}.
They prove that the solution to the adjoint problem with their boundary
condition gives the PRC for the asymptotic phase of hybrid systems.
Special cases of this result have also been presented by Ladenbauer
et al. \citep{ladenbauer2012ImpactAdaptationCurrents} for a specific
hybrid neuron model and by Park et al. \citep{park2016InfinitesimalPhaseResponse}
for nonsmooth systems with discontinuous boundaries but no reset map.
We present a brief heuristic derivation for an equivalent adjoint
boundary condition that follows directly from the existence of an
appropriate differentiable phase mapping, and show that the condition
has an intuitive form tied to the geometry of the threshold. 

\begin{figure}
\begin{centering}
\includegraphics[scale=0.7]{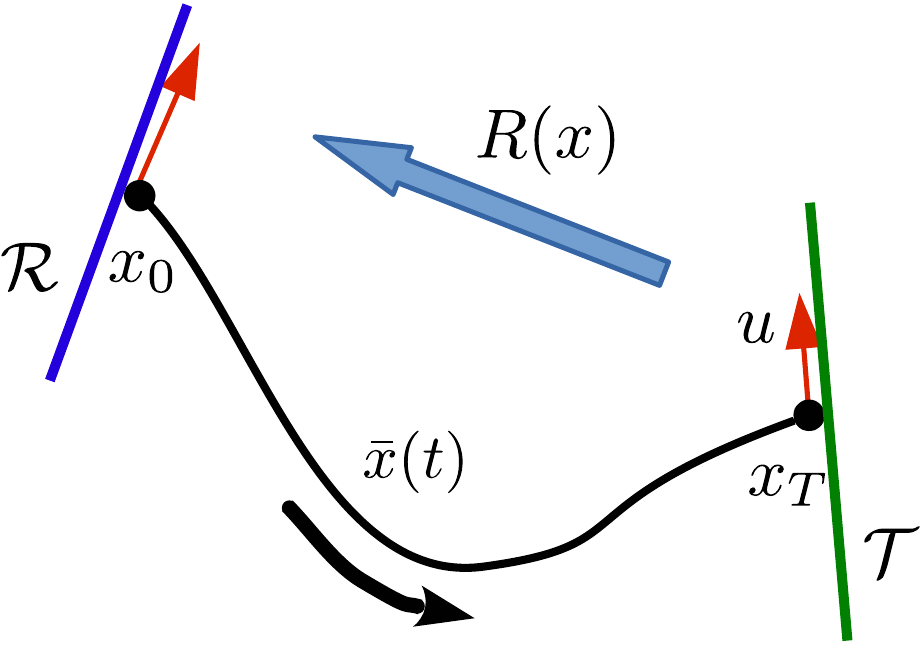}
\par\end{centering}
\caption{\label{fig:reset_drawn}Limit cycle trajectory $\bar{x}\left(t\right)$
evolves smoothly from initial condition $x_{0}$ on the reset manifold
$\mathcal{R}$ to $x_{T}$ on the threshold manifold $\mathcal{T}$,
then is returned by the reset map $R\left(x\right)$ to $x_{0}=R\left(x_{T}\right)$.
A perturbation $u$ along the threshold is mapped to a post-reset
perturbation along the reset manifold.}
\end{figure}

Consider a limit cycle trajectory $\bar{x}\left(t\right)$, reaching
the threshold manifold at $\bar{x}\left(T\right)=x_{T}=\left(v_{T},\,w_{T}\right)$,
as shown in \figref{reset_drawn}. The phase on this trajectory cannot
change across the instantaneous reset. That is, since $\Theta\left(\bar{x}\left(t\right)\right)$
must be continuous in $t$, 
\begin{equation}
\Theta\left(x_{T}\right)=\lim_{t\to T^{-}}\Theta\left(\bar{x}\left(t\right)\right)=\lim_{t\to T^{+}}\Theta\left(\bar{x}\left(t\right)\right)=\Theta\left(R\left(x_{T}\right)\right),\label{eq:phase_cont}
\end{equation}
where the right limit of the reset discontinuity is given by the reset
point $R\left(x_{T}\right)$. We introduce a unit tangent vector $u$
along the threshold manifold $\mathcal{T}$ at $x_{T}$ (\figref{reset_drawn}).
Assuming the PRC is well-defined before and after reset, we can apply
a directional derivative along this tangent vector ($D_{u}$) to both
sides of (\ref{eq:phase_cont}). 
\begin{eqnarray}
D_{u}\Theta\left(x_{T}\right) & = & D_{u}\Theta\left(R\left(x_{T}\right)\right),\nonumber \\
\nabla\Theta\left(x_{T}\right)^{T}u & = & \nabla\Theta\left(R\left(x_{T}\right)\right)^{T}D_{u}R\left(x_{T}\right),\nonumber \\
Z_{u}\left(T^{-}\right) & = & \left(D_{u}R\left(x_{T}\right)\right)^{T}Z\left(0^{+}\right).\label{eq:boundary}
\end{eqnarray}
This equation gives a boundary condition for the PRC, replacing the
standard assumption of periodicity. Together with the normalization
condition (\ref{eq:normalize}), it determines the unique solution
to the adjoint problem for the discontinuous limit cycle of the hybrid
model. With the reset map only defined on the threshold manifold,
the derivative is only defined in the tangent direction, and there
is no corresponding constraint on the perpendicular component of the
PRC. In $N$ dimensions the threshold manifold is $\left(N-1\right)$-dimensional,
so rather than a single vector $u$, we enforce (\ref{eq:boundary})
for each of $N-1$ vectors $u_{i}$ spanning the tangent space.\footnote{By extending the reset map to a neighborhood of the boundary, Shirasaka
et al. \citep{shirasaka2017Phasereductiontheory} instead present
$N$ conditions; the $N$th condition missing from our analysis is
redundant if the normalization condition (\ref{eq:normalize}) is
enforced at all times (see \subsecref{Connection-to-shira}).}

This boundary condition expresses an intuitive fact about perturbations
to the limit cycle: the phase difference between the limit cycle and
a trajectory perturbed along the threshold (given by $Z_{u}\left(T^{-}\right)$)
must be the same as the difference after both trajectories are reset.
The difference after reset is expressed by the PRC at the reset point,
$Z\left(0^{+}\right)$, with the perturbation transformed by the reset
map to a distinct perturbation along the reset manifold, approximated
by $D_{u}R$ (\figref{reset_drawn}).

\section{Phase reduction of the resonate-and-fire model\label{sec:Resonate-and-fire-phase-reductio}}

\subsection{Existence and stability of spiking limit cycles\label{subsec:spiking-limit-cycles}}

Before applying the theory of phase reduction to any model, we must
ensure the system exhibits a stable limit cycle. We begin our analysis
of the resonate-and-fire model by finding the existence and stability
conditions for spiking limit cycles. These conditions define boundaries
of the stable limit cycle regime in parameter space, bifurcations
of the model.

\subsubsection*{Hard reset}

We first discuss the existence conditions of the spiking limit cycle
with hard reset, (\ref{eq:resfire_reset}). The spiking limit cycle
exists whenever a trajectory starting from the reset point crosses
the threshold. The limit cycle is lost in a ``grazing bifurcation''
when the trajectory becomes tangent to and then fails to cross the
firing threshold. Beyond this bifurcation, trajectories show decaying
subthreshold oscillations, approaching rest at the equilibrium voltage.
The hard reset map ensures that the spiking limit cycle is always
stable, as the reset erases all effects of small perturbations to
the cycle by projecting to the single reset point. 

\begin{figure}
\begin{centering}
\includegraphics{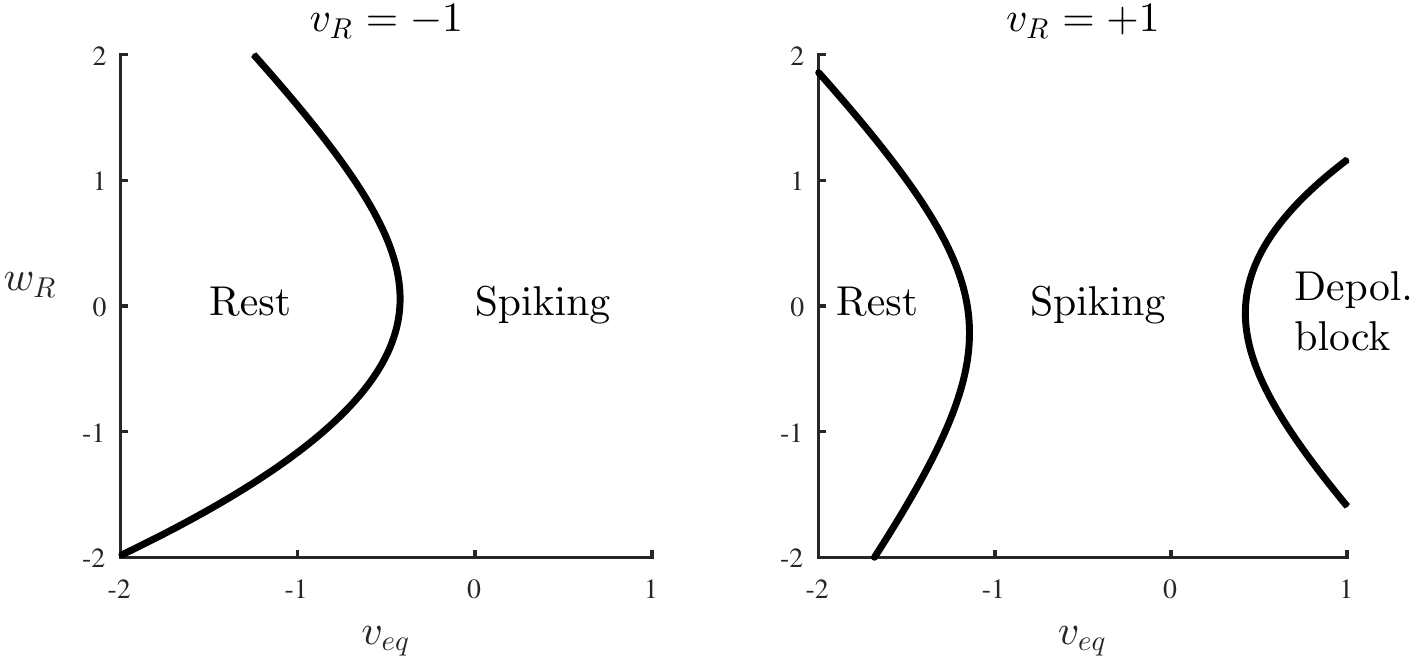}
\par\end{centering}
\caption{\label{fig:spike-paramspace}Grazing bifurcations bounding the spiking
regime in parameter space, for positive and negative reset regimes
of the hard reset resonate-and-fire model with \textcolor{black}{$\lambda=0.1$.
In both rest and depolarization block regimes, the system has a stable
fixed point (quiescent state); this state is below threshold at rest
and above threshold for depolarization block.}}
\end{figure}

We can visualize the spiking regime boundaries in a two-dimensional
parameter space that captures the most important dimensions of variability
of the model dynamics. We fix the frequency $\omega=1$ without loss
of generality, and choose a small decay parameter $\lambda=0.1$ to
give slowly decaying subthreshold oscillations. The dynamics then
depend on the reset parameters $v_{R}$ and $w_{R}$ as well as the
equilibrium $v_{eq}$, but because the model is invariant to uniform
rescaling of the ($v$, $w$) phase space, we can rescale to $\left|v_{R}\right|=1$
without loss of generality. Therefore, in \figref{spike-paramspace}
we explore two distinct two-dimensional ($v_{eq}$, $w_{R}$) parameter
spaces for positive and negative reset voltage, fixed at $v_{R}=\pm1$
(see \figref{rf_traj}). In the both negative and positive reset regimes,
a single grazing bifurcation occurs for low equilibrium voltage $v_{eq}$
(negative tonic input current), below which the model is quiescent
(at rest). In the positive reset regime, a second grazing bifurcation
occurs for high $v_{eq}$, when the voltage fails to dip below threshold,
corresponding to depolarization block. The bifurcations shown in \figref{spike-paramspace}
curve away from the origin because any increase in the magnitude of
$w_{R}$ increases the radius of the orbit, moving the system away
from the bifurcation. We can express the tangency condition for the
grazing bifurcation in terms of the orientation of the trajectory
when crossing threshold, $\theta_{H}=\tan^{-1}\left(\frac{\dot{v}\left(T\right)}{\dot{w}\left(T\right)}\right)$.
At the bifurcation, this orientation matches that of the threshold,
\begin{equation}
\theta_{H}=\theta_{0}+T+\frac{\pi}{2}+\tan^{-1}\lambda=\pm\frac{\pi}{2}.\label{eq:grazing}
\end{equation}

\subsubsection*{Soft reset}

Although the subthreshold dynamics always lead to decay of perturbations,
the soft reset map can in some cases amplify perturbations, making
the limit cycle unstable. In addition to the grazing bifurcation boundaries,
the stable spiking regime can also be lost in a saddle-node bifurcation
of limit cycles, where the stable and unstable cycles collide and
annihilate. In addition, the dynamics with soft reset allow for limit
cycles with multiple spikes, and we will show that the spiking limit
cycle can also lose stability in a period-doubling bifurcation.

To determine existence and stability conditions for the soft reset
limit cycles, we reduce the dynamics to a Poincaré return map on the
reset manifold (also referred to as an adaptation map \citep{touboul2009SpikingDynamicsBidimensional}).
This map takes a value of the adaptation variable at the $k^{\mathrm{th}}$
reset, $w_{k}$, to the value at the following reset, $w_{k+1}=P\left(w_{k}\right)$.
A limit cycle corresponds to a fixed point of the map, $\bar{w}=P\left(\bar{w}\right)$,
and the cycle is asymptotically stable if the fixed point satisfies
$\left|\frac{dP}{dw}\left(\bar{w}\right)\right|<1.$ The derivative
of the return map characterizes the degree of attraction to the limit
cycle and is therefore also tied to the validity of the asymptotic
approximation of the phase reduction. We begin by deriving the return
map for a general reset map from the threshold $\left(v_{T},\,w\right)$
to $R\left(w\right)=\left(v_{R},\,R_{w}\left(w\right)\right)$. We
then evaluate the fixed points and their stability for the soft reset
map. 

We define the flow $F\left(w_{k},\,t\right)=x\left(t\right)=\left(v\left(t\right),\,w\left(t\right)\right)^{T}$,
where the trajectory $x\left(t\right)$ satisfies the subthreshold
dynamics from \eqref{resfire_linear} starting from an initial condition
$x\left(0\right)=\left(v_{R},\,w_{k}\right)^{T}$ on the reset manifold.
The flow evaluates to
\begin{equation}
F\left(w_{k},\,t\right)=\left(\begin{array}{c}
F_{v}\\
F_{w}
\end{array}\right)=e^{-\lambda t}\left(\begin{array}{cc}
\cos t & -\sin t\\
\sin t & \cos t
\end{array}\right)\left(\begin{array}{c}
v_{R}\\
w_{k}
\end{array}\right).\label{eq:flow}
\end{equation}
We then define the spike time map $\tau\left(w_{k}\right)$, giving
the time it takes such a trajectory to reach the threshold,
\begin{equation}
\tau\left(w_{k}\right)=\min\left\{ t:\ F_{v}\left(w_{k},\,t\right)=v_{T}\right\} .\label{eq:spike time map}
\end{equation}
A trajectory starting at $w_{k}$ crosses threshold at the point $F\left(w_{k},\,\tau\left(w_{k}\right)\right)$,
and is reset to $R\left(F_{w}\left(w_{k},\,\tau\left(w_{k}\right)\right)\right)$.
The $w$-component of this reset point, $R_{w}$, gives the desired
return map,

\begin{equation}
w_{k+1}=P\left(w_{k}\right)=R_{w}\left(F_{w}\left(w_{k},\,\tau\left(w_{k}\right)\right)\right).\label{eq:adaptation map}
\end{equation}

A fixed point $\bar{w}=w_{0}$ of the return map corresponds to a
limit cycle with period $T=\tau\left(w_{0}\right)$. The stability
of a limit cycle with soft reset is assessed by evaluating the derivative
of the return map for the soft reset rule $R_{w}\left(w\right)=w+\Delta w$
(\ref{eq:softreset}).

\begin{eqnarray*}
\frac{dP}{dw}\left(w_{0}\right) & = & \frac{dR_{w}}{dw}\left(w_{T}\right)\left(\frac{\partial F_{w}}{\partial w}\left(w_{0},\,T\right)+\frac{\partial F_{w}}{\partial t}\left(w_{0},\,T\right)\frac{\partial\tau}{\partial w}\left(w_{0}\right)\right)\\
 & = & \left(\frac{\partial F_{w}}{\partial w}\left(w,\,T\right)-\frac{dw}{dt}\left(T\right)\frac{\frac{\partial F_{v}}{\partial w}\left(w_{0},\,T\right)}{\frac{dv}{dt}\left(T\right)}\right)\\
 & = & e^{-\lambda T}\left(\cos T+\tan\theta_{H}\sin T\right),
\end{eqnarray*}
where we recall the trajectory's orientation at threshold $\theta_{H}=\theta_{0}+T+\frac{\pi}{2}+\tan^{-1}\lambda$.
For the soft reset rule then, the limit cycle can be stable or unstable
depending on the decay $\lambda$ and the geometry of the threshold
and reset manifolds. Loss of stability occurs when 
\[
\pm1=\frac{dP}{dw}\left(w_{0}\right)=e^{-\lambda T}\left(\cos T+\tan\theta_{H}\sin T\right),
\]
which implies that bifurcations occur in the full resonate and fire
model when 
\begin{gather}
\tan\theta_{H}=\frac{\pm e^{\lambda T}-\cos T}{\sin T}.\label{eq:cycle_instability}
\end{gather}
The negative slope instability corresponds to a subcritical (proof
not shown) period doubling bifurcation, where an unstable period-two
limit cycle collides with a stable period-one limit cycle. The positive
slope instability corresponds to a saddle-node bifurcation of limit
cycles, with a stable cycle coalescing either with a finite unstable
cycle, or at infinity (return map slope approaching unity as $w_{0}\to-\infty$
for finite parameter values). 

\begin{figure}
\begin{centering}
\includegraphics[scale=0.8]{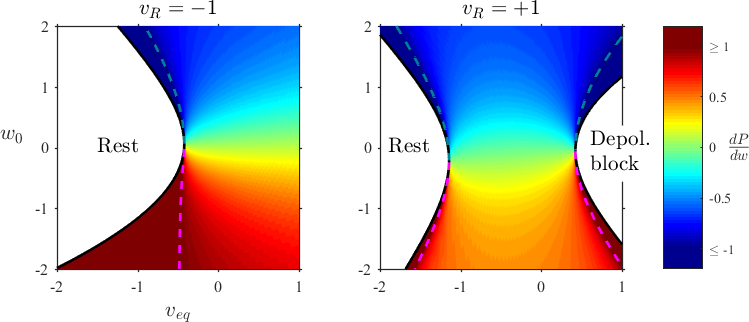}
\par\end{centering}
\caption{\label{fig:soft-paramspace}Existence and stability of limit cycles
for soft reset, for the negative reset (left) and positive reset (right)
regimes. Color gives the stability quantified as the derivative of
the return map $P$. Dashed lines show loss of stability, with blue-green
the negative slope (period-doubling) and magenta the positive slope
(saddle-node) bifurcation. Solid red/blue indicates unstable cycles
past the bifurcation, slopes $\left|\frac{dP}{dw}\right|>1$. Black
lines show the grazing bifurcation of the unstable limit cycle. Parameters
$\lambda=0.1$ and $v_{R}=\pm1$ fixed while varying $v_{eq}$ and
$\Delta w$, plotted using $w_{0}$ to facilitate comparison with
hard reset parameter space.}
\end{figure}

In \figref{soft-paramspace}, we plot the stability and grazing bifurcations
for soft reset together in $\left(v_{eq},\,w_{0}\right)$ parameter
space. Since $w_{0}$ corresponds to the parameter $w_{R}$ for hard
reset, the grazing bifurcations (solid lines in \figref{soft-paramspace})
are identical to the hard reset grazing bifurcations in $\left(v_{eq},\,w_{R}\right)$
coordinates from \figref{spike-paramspace}. The loss of stability
bifurcations and the grazing bifurcations form the two boundaries
of a narrow unstable limit cycle regime. These bifurcations are both
related to threshold crossing and necessarily lie close together.
Perturbations are amplified by the reset map, causing instability,
because of a mismatch between the angles of incidence with the threshold
and reset manifolds, which increases as the trajectory approaches
tangency to the threshold (the grazing bifurcation condition). We
note that, near the saddle-node bifurcations, the return map can have
two fixed points, representing stable and unstable cycles. The representation
in \figref{soft-paramspace} unfolds the bifurcation so that a single
point in the parameter space $\left(v_{eq},\,\Delta w\right)$ corresponds
to two points in $\left(v_{eq},\,w_{0}\right)$ space, stable and
unstable cycles on opposite sides of the dashed bifurcation line.

\subsection{\label{subsec:RF-PRC}Resonate-and-fire PRC}

The first step in proceeding with the phase reduction of the resonate-and-fire
model is to evaluate its phase response curve (PRC), expressing the
effect of perturbations to the limit cycle as a proportional phase
shift (\subsecref{Phase-response-curve}). In general, the PRC can
be evaluated either by direct calculation or by the adjoint method.
Although directly calculating the effect of perturbations typically
must be carried out computationally, the hard reset of the original
resonate-and-fire model allows the PRC to be calculated directly from
an analytical expression for phase \citep{miura2006Globallycoupledresonateandfire}.
Since all points on the threshold are mapped onto the single point
$x_{R}$, the threshold itself is an isochron and can serve as a reference
point to define the phase map for all trajectories. However, we take
a different approach deriving the PRC from the adjoint equation for
a general reset rule, following the theory described in \subsecref{Hybrid-PRC}.
This approach allows us to see the hard reset as a special case and
to capture the geometric intuition in the relationship between the
dynamics, the adjoint equation, and the reset rule. 

\begin{figure}
\centering{}\includegraphics{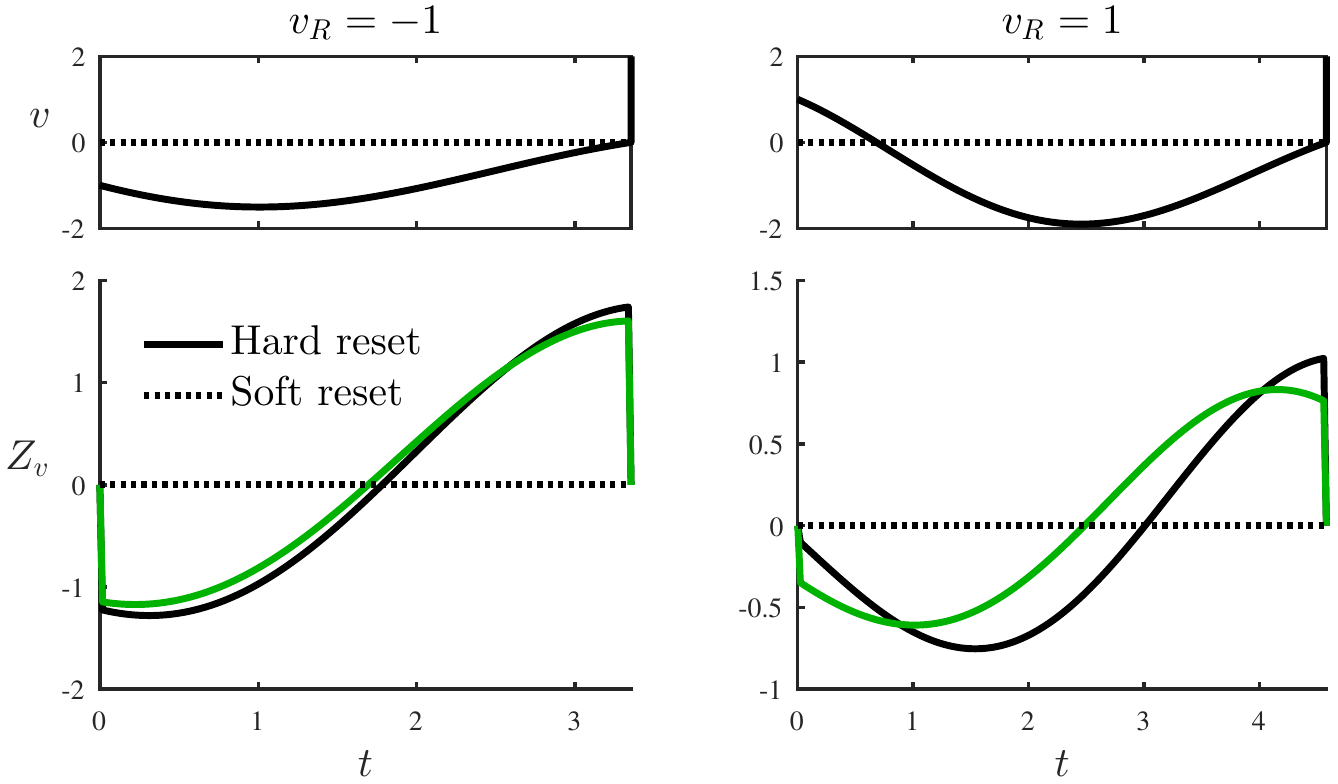}\caption{\label{fig:rf_z}The $v$-component PRC $Z_{v}\left(t\right)$, derived
by the adjoint method for both soft and hard reset. Parameters: $\lambda=0.1,\,v_{R}=1$,
$w_{R}=w_{0}=1$, $v_{eq}=-0.5$. For soft reset, $\Delta w$ was
set to match the hard reset limit cycle, so that $w_{0}=w_{R}$.}
\end{figure}

We proceed by first evaluating the adjoint equation (\ref{eq:adjoint})
for the subthreshold dynamics. In general, the adjoint equation evaluates
the dynamics linearized about the limit cycle. Since the resonate-and-fire
dynamics (\ref{eq:resfire_linear}) are linear, the adjoint equation
is simply defined by the negative transpose of the time-independent
linear operator,
\begin{equation}
\frac{dZ}{dt}=-Df\left(\bar{x}\right)^{T}Z=\begin{pmatrix}\lambda & -1\\
1 & \lambda
\end{pmatrix}Z.\label{eq:resfire adjoint}
\end{equation}
The PRC solution exhibits exponentially growing oscillations,
\begin{equation}
\begin{alignedat}{1}Z_{v}\left(t\right) & =\frac{A}{r_{0}}e^{\lambda t}\cos\left(t-T+\alpha\right),\\
Z_{w}\left(t\right) & =\frac{A}{r_{0}}e^{\lambda t}\sin\left(t-T+\alpha\right).
\end{alignedat}
\label{eq:resfire prc}
\end{equation}
The PRC is defined as written for times $0<t<T$, and extends periodically
to all $t$ modulo $T$, with a discontinuity at $t=0$ (due to the
discontinuous reset map skipping over dynamics during the spike).
The amplitude $A=\frac{1}{\sqrt{1+\lambda^{2}}\cos\left(\theta_{H}-\alpha\right)}$
is determined by the normalization condition $Z\left(T\right)\cdot\frac{dx}{dt}\left(T\right)=1$.
Based on the reset map, the boundary condition for the adjoint equation
(\ref{eq:boundary}) links the left and right limits of the discontinuity,
determining the phase shift $\alpha$. Examples of the $v$-component
PRC for both positive and negative $v_{R}$ are shown in \figref{rf_z}.

The general form for the boundary condition from (\ref{eq:boundary})
simplifies given our assumption that the threshold and reset manifolds
are in the $w$-direction (constant $v$), with reset map $R_{w}\left(w\right)$
\begin{equation}
Z_{w}\left(T^{-}\right)=\frac{dR_{w}}{dw}\left(w_{T}\right)Z_{w}\left(0^{+}\right).\label{eq:bc_thresh}
\end{equation}
For the hard reset (\ref{eq:resfire_reset}), mapping to a constant
reset point, the derivative of the reset map is zero. Thus (\ref{eq:bc_thresh})
reduces to the terminal condition $Z_{w}\left(T^{-}\right)=0$, corresponding
to a phase $\alpha=0$. This result is equivalent to the geometric
constraint that the PRC must be perpendicular to the threshold, or
oriented in the $v$-direction at time $T$. That is, perturbations
along the threshold have no effect after the reset, and the threshold
is an isochron, as shown in \figref{rf_pp}.

For the soft reset rule (\ref{eq:softreset}), an increment of $w$,
the boundary condition (\ref{eq:bc_thresh}) mandates continuity of
the $w$-component of the PRC, $Z_{w}\left(T^{-}\right)=Z_{w}\left(0^{+}\right).$
Intuitively, this tells us that a perturbation to $w$ immediately
before the spike has the same effect as a perturbation after the spike;
i.e., perturbations tangent to the threshold are unchanged by the
soft reset map. This continuity boundary condition leads to a phase
shift of $Z_{v}$,
\[
\alpha=\arctan\left(\frac{\sin T}{\cos T-e^{\lambda T}}\right).
\]
Note that this implies that the threshold manifold is not an isochron.
This phase shift is typically small for positive $v_{R}$ but can
grow more significant in parts of the negative reset regime. (See
example in \figref{rf_z}, and full calculation in \subsecref{PRC-phase}.)

We note that in the more general case of a nonlinear reset map, the
result depends on $\frac{dR_{w}}{dw}$, for which the hard and soft
reset are the special cases $\frac{dR_{w}}{dw}=0$ and 1 respectively.
A nonlinear reset map with derivative close to either extreme would
lead to small corrections to the corresponding $\alpha$ value. Similarly,
small variations in the orientation of the threshold or reset manifold
lead to minor adjustments to (\ref{eq:bc_thresh}) and to the resulting
phase shift $\alpha$.

We also note that the amplitude $A$ has a singularity when 
\begin{equation}
\cos\left(\theta_{H}-\alpha\right)=0.\label{eq:singularity}
\end{equation}
This singularity is associated with the high sensitivity to perturbations
near bifurcations of the limit cycle. For the hard reset $\alpha=0$,
$\theta_{H}=\pm\frac{\pi}{2}$ is the grazing bifurcation condition
(\ref{eq:grazing}). For the soft reset, the singularity occurs when
$\tan\theta_{H}=-\cot\alpha=\frac{e^{\lambda T}-\cos T}{\sin T}$,
which is condition (\ref{eq:cycle_instability}) for the saddle-node
bifurcation of the limit cycle (positive slope instability of the
return map). Interestingly, the negative slope loss of stability (period-doubling
bifurcation) is not reflected in the PRC. It seems that in the positive
slope instability, phase shifts continue to accumulate progressively
on each cycle, while in the negative slope case, positive and negative
phase shifts alternate as they grow, leading the PRC to reflect an
averaged finite phase shift that fails to capture the loss of stability.

\subsection{Resonate-and-fire phase model}

We now proceed to construct the interaction function and phase model
for electrically-coupled resonate-and-fire neurons. In the weak coupling
regime, this reduced model with a single phase variable for the state
of each oscillator captures the full synchronization properties of
the resonate-and-fire network (\ref{eq:resfire net}). The interaction
function $H$ and heterogeneity $\Omega$ of the phase model are calculated
according to equation (\ref{eq:phase_model}), from the electrical
coupling and heterogeneity of frequencies in the full model along
with the PRC from (\ref{eq:resfire prc}).

The heterogeneity of frequencies in the phase model follows directly
from the resonate-and-fire model's frequency heterogeneity. Evaluating
the integral (\ref{eq:phase_het}) and applying the normalization
condition (\ref{eq:normalize}) shows that $\Omega_{i}=\omega_{i}$
(the subthreshold angular frequency). We note that weak heterogeneity
in any other parameter of the resonate-and-fire model (e.g., $v_{R}$)
would create an additional additive contribution to the phase model
heterogeneity $\Omega_{i}$ proportional to that parameter heterogeneity. 

\begin{figure}
\begin{centering}
\includegraphics{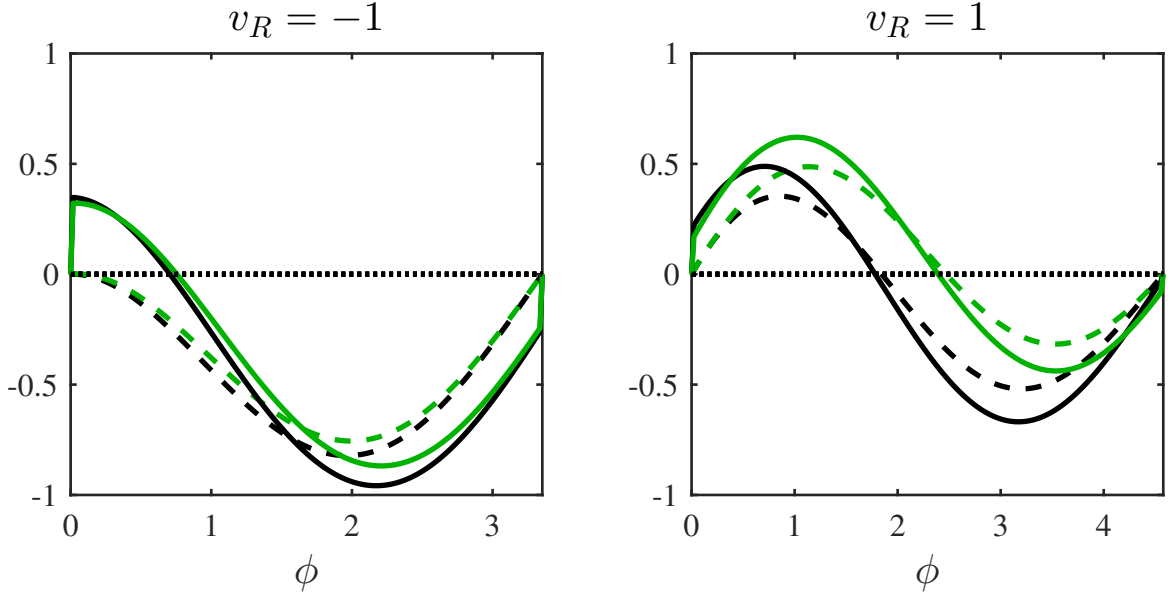}
\par\end{centering}
\caption{\label{fig:rf_h}Interaction function $H$ including both spike and
subthreshold contributions (solid lines), as well as subthreshold
component $H_{sub}$ (dashed), for the electrically coupled resonate-and-fire
model with hard reset (black) and soft reset (green). Parameters:
$\lambda=0.1,\,w_{R}=1,\,v_{eq}=-0.5,\,M=0.2$.}
\end{figure}

The interaction function $H$ defines the nonlinear coupling between
cells in the phase model. It can be expressed as a convolution integral
of the coupling current and the PRC, according to (\ref{eq:phase_interaction}).
In the case of electrical coupling, the coupling current $I_{c}=v\left(t+\phi\right)-v\left(t\right)$
depends on the voltage component of the limit cycle. The resulting
interaction function, depicted in \figref{rf_h}, is 
\begin{align}
H\left(\text{\ensuremath{\phi}}\right) & =\frac{1}{T}\int_{0}^{T}Z_{v}\left(t\right)\left[v\left(t+\phi\right)-v\left(t\right)\right]dt.
\end{align}
Since the interaction function convolves the PRC and the voltage limit
cycle, with a linear dependence on both, we can separate the subthreshold
and spiking components, $H=H_{sub}+H_{spike}$, corresponding to the
subthreshold and spiking components of the limit cycle from equation
(\ref{eq:resfire trajectory}), $\bar{v}\left(t\right)=v_{sub}(t)+M\delta\left(t\right)$. 

\begin{eqnarray*}
H_{sub}\left(\phi\right) & = & \frac{1}{T}\int_{0}^{T}Z_{v}\left(t\right)\left[v_{sub}\left(t+\phi\right)-v_{sub}\left(t\right)\right]dt,\\
H_{spike}\left(\phi\right) & = & \frac{M}{T}\int_{0}^{T}Z_{v}\left(t\right)\left[\delta\left(t+\phi\right)-\delta\left(t\right)\right]dt.
\end{eqnarray*}

The spike $\delta$-function component of the limit cycle determines
the \emph{spike interaction function} $H_{spike}$, representing the
effect of this voltage transient through the electrical coupling.
This is essentially a pulse-coupling interaction, as used in simple
models of excitatory chemical synapses \citep{mirollo1990SynchronizationPulseCoupledBiological,kuramoto1991Collectivesynchronizationpulsecoupled,miura2004Pulsecoupledresonateandfiremodels},
and the effect is entirely determined by the PRC and the amplitude
of the spike. With zero-width or $\delta$-function pulses, the spike
interaction function can be discontinuous at the origin, as shown
in \figref{rf_h}. We return to the effects of this interaction in
\subsecref{Spike-phase-interaction}.

The \emph{subthreshold interaction function} $H_{sub}$ captures the
effect of subthreshold fluctuations of the limit cycle, $v_{sub}(t)=r_{0}e^{-\lambda t}\cos\left(t+\theta_{0}\right)$.
Analysis of this component will allow us to determine how the resonant
properties of the model contribute to synchronization. Each parameter
of the model affects synchronization both through its effect on the
limit cycle and on the PRC, making the combined effect (encoded by
$H_{sub}$) potentially complex. The only general constraints on the
subthreshold interaction function are that it must be continuous and
pass through the origin; $H_{sub}\left(0\right)=0$ because the gap
junction coupling is diffusive, proportional to the difference of
voltages.

\begin{figure}
\begin{centering}
\includegraphics[scale=0.6]{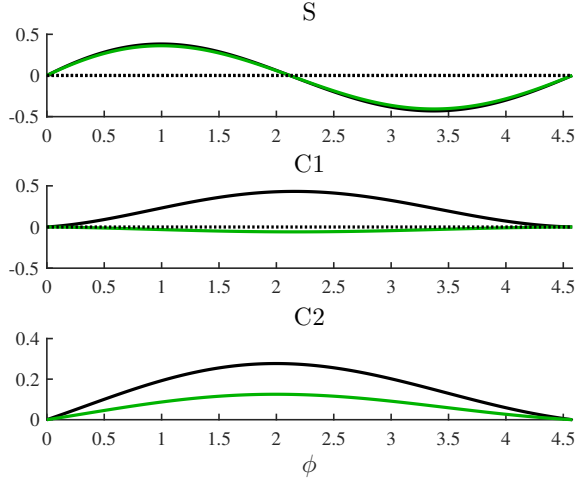}
\par\end{centering}
\caption{\label{fig:rf_Hcomps}Components of the subthreshold interaction function
$H_{sub}$ with approximate odd and even symmetry, for the resonate-and-fire
model with hard reset (black) and soft reset (green). Parameters:
$\lambda=0.1,\,v_{R}=w_{R}=1,\,v_{eq}=-0.5$.}
\end{figure}

To simplify the analysis, we split the subthreshold interaction function
into three terms with distinct parameter dependence (calculation in
\subsecref{app-Subthreshold-interaction-functio}). 

\begin{equation}
\begin{alignedat}{1}H_{sub}(\phi) & =A_{C1}C1(\phi)+A_{C2}C2(\phi)+A_{S}S(\phi),\\
C1(\phi) & =1-\frac{e^{-\lambda\phi}}{T}\left[e^{\lambda T}\phi\cos\left(T-\phi\right)+\left(T-\phi\right)\cos\phi\right],\\
C2(\phi) & =\frac{e^{-\lambda\phi}}{T}\left[e^{\lambda T}\sin\phi+\sin\left(T-\phi\right)\right]-\frac{\sin T}{T},\\
S(\phi) & =\frac{e^{-\lambda\phi}}{T}\left[-e^{\lambda T}\phi\sin\left(T-\phi\right)+\left(T-\phi\right)\sin\phi\right].
\end{alignedat}
\label{eq:H comps}
\end{equation}
\begin{equation}
A_{C1}=-\frac{A}{2}\cos\left(\theta_{T}-\alpha\right),\,A_{C2}=\frac{A}{2}\cos\left(\theta_{0}+\alpha\right),\,A_{S}=-\frac{A}{2}\sin\left(\theta_{T}-\alpha\right),\label{eq:H coeff}
\end{equation}
where $\theta_{T}=\theta_{0}+T$ is the angular coordinate of the
trajectory at threshold. We note that the $S$-function closely resembles
sine, while the two $C$-functions resemble a vertically shifted cosine,
as shown in the example in \figref{rf_Hcomps}. This approximate odd
and even symmetry about the origin means they contribute in distinct
ways to synchronization of simple networks. In the following sections,
we consider these effects by studying networks of two and three cells.

\section{Synchronization of two electrically coupled resonate-and-fire neurons\label{sec:odd-pair}}

Our primary goal with the phase reduction of the resonate-and-fire
model is to provide insight into the synchronization of networks of
electrically coupled resonant neurons. Even after phase reduction,
the analysis of large systems with realistic network architecture
is hindered by the nonlinear phase coupling and number of degrees
of freedom, and in general must be carried out numerically. Therefore,
we focus on minimal networks of two or three cells and save the analysis
of large-scale networks for future work. In this idealized context,
we can explain how the cellular properties of the resonate-and-fire
oscillators determine network synchrony through the distinct components
of the interaction function (in terms of slopes, amplitudes and discontinuities).
Although large networks cannot be completely understood by their two-
and three-cell subnetworks, in many cases the intuition built on these
minimal networks holds \citep{acebron2005Kuramotomodelsimple}.

\subsection{General considerations}

We begin by examining the synchronization of a symmetrically coupled
pair of oscillators, the simplest and most commonly analyzed network
\citep{ermentrout1981Phaselockingweaklycoupled,kopell2002Mechanismsphaselockingfrequency,lewis2003Dynamicsspikingneurons,mancilla2007SynchronizationElectricallyCoupled}.
The assumption of symmetry implies that the equation governing the
phase difference between the oscillators (\ref{eq:pair}) isolates
a component of the interaction function with odd symmetry, simplifying
the analysis. Experiments have shown that electrical coupling is often
close to symmetric, especially when between the same cell type and
formed by the common symmetric channel type connexin-36 \citep{bennett2004ElectricalCouplingNeuronal}.
Nonetheless, asymmetry can and does arise, either from rectification
(favoring current flow in one direction) in the gap junction channels
that make up the electrical synapse, or from asymmetries in size or
gap junction location. We will address the effects of asymmetry briefly
in \secref{Even-3}.

\begin{figure}
\begin{tabular}{>{\centering}m{0.5\textwidth}>{\centering}m{0.4\linewidth}}
\subfloat[\label{fig:pair-example-Hodd}]{\begin{centering}
\includegraphics[width=0.45\textwidth]{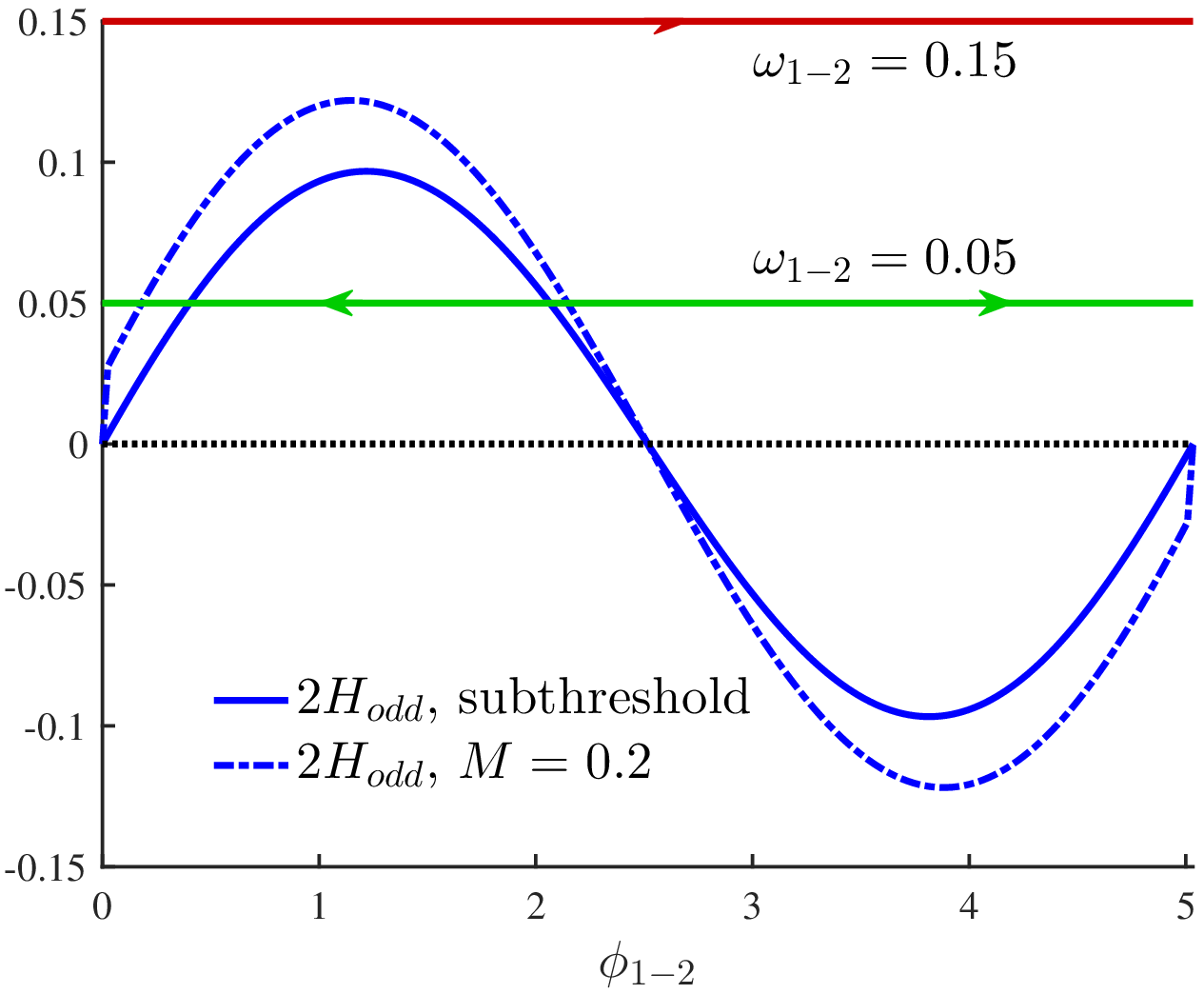}
\par\end{centering}
} & \subfloat[\label{fig:pair-spikes}]{\begin{centering}
\begin{tabular}{c}
\includegraphics[width=0.35\textwidth]{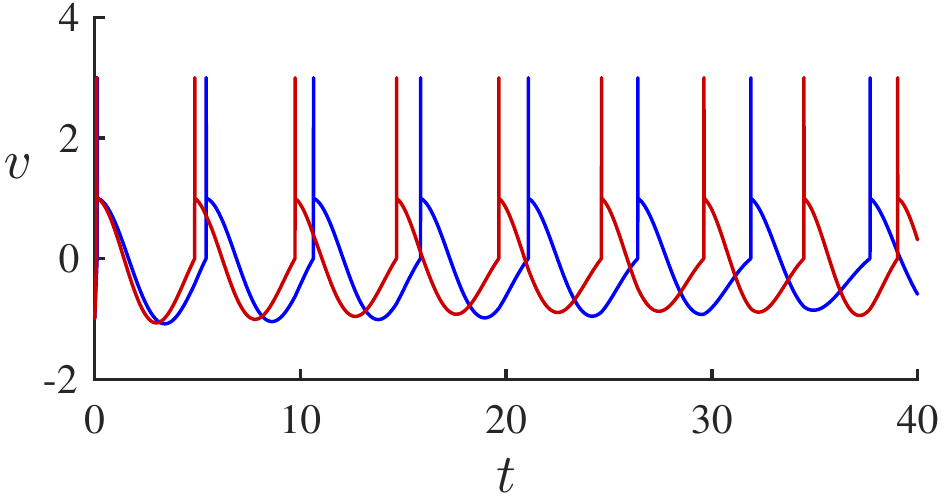}\tabularnewline
\includegraphics[width=0.35\textwidth]{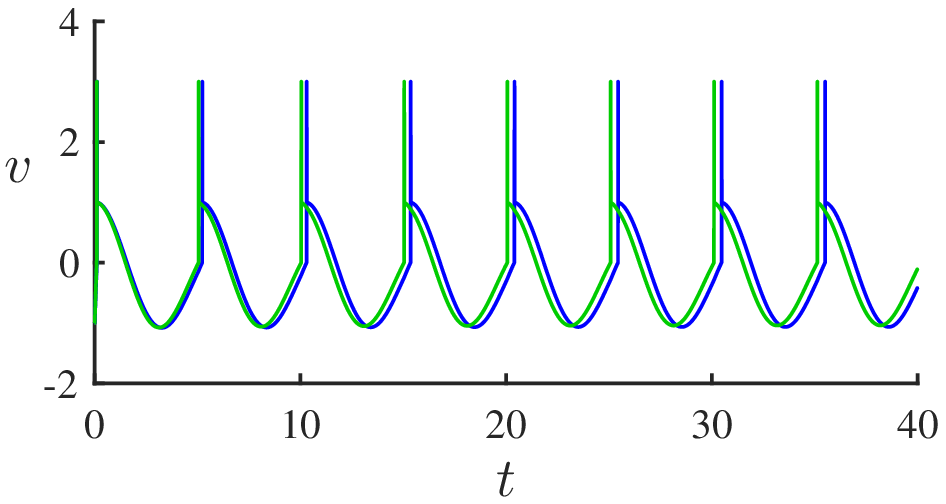}\tabularnewline
\end{tabular}
\par\end{centering}
}\tabularnewline
\end{tabular}

\caption{\label{fig:pair-example}Phase locking of a coupled pair. (a): Phase-locked
states are fixed points of (\ref{eq:pair}), intersections of $2H_{odd}\left(\phi\right)$
(blue) with lines of constant frequency heterogeneity $\omega_{1-2}$
(red/green). Arrows give the flow of relative phase (\ref{eq:pair}),
indicating existence and stability of fixed points. The spiking component
of $H_{odd}$ provides additional robustness (dashed line with spike,
solid line without). (b): Simulations show phase-locked (green) and
phase walk-through (red) states corresponding to two levels of heterogeneity
from A (subthreshold coupling only). Parameters: hard reset, $\lambda=0.1,\,w_{R}=1$,
$v_{R}=1$, $v_{eq}=-0.5,\,M=0$, $K=0.1$.}
\end{figure}

The phase model (\ref{eq:phase_model}) for a pair of symmetrically
coupled oscillators ($k_{12}=k_{21}=K$) is given by
\begin{equation}
\dot{\theta}_{i}=\omega_{i}+KH\left(\theta_{j}-\theta_{i}\right).\label{eq:pair-each}
\end{equation}
Expressed in terms of the phase difference $\phi_{i-j}=\theta_{i}-\theta_{j}$
and frequency difference $\omega_{1-2}$, (\ref{eq:pair-each}) reduces
to
\begin{equation}
\dot{\phi}_{1-2}=\omega_{1-2}-2KH_{odd}\left(\phi_{1-2}\right),\label{eq:pair}
\end{equation}
 where $H_{odd}\left(\phi\right)=\nicefrac{1}{2}\left(H\left(\phi\right)-H\left(-\phi\right)\right)=H\left(\phi\right)-H_{even}\left(\phi\right)$.
(Some analyses refer to the $G$-function, $G\left(\phi\right)=-2H_{odd}\left(\phi\right)$
\citep{schwemmer2012TheoryWeaklyCoupled}). Fixed points of (\ref{eq:pair})
correspond to phase-locked states of the coupled pair, the existence
and stability of which are determined by properties of $H_{odd}$,
as depicted in \figref{pair-example-Hodd} and described below.

For identical oscillators ($\omega_{1-2}=0$), the odd symmetry of
\eqref{pair} implies a pair of fixed points at $\phi_{1-2}=0$ and
$\phi_{1-2}=\nicefrac{T}{2}$, for synchronous and antiphase states
respectively. If $H_{odd}$ has only a single local maximum, which
is typical for the resonate-and-fire model, only one of these two
fixed points is stable and no additional fixed points exist. The synchronous
state is stable and the antiphase state unstable when the slope $H_{odd}^{\prime}\left(0\right)$
is positive, and the reverse holds for negative slope. We note that
in larger networks a similar dependence on the slope can be shown:
a strong positive slope leads to global synchrony, while a negative
slope leads to global incoherence \citep{acebron2005Kuramotomodelsimple}.
(For simplicity, we shorten references to the slope evaluated at the
origin to ``slope''.) For hybrid models, the slope of the full interaction
function $H^{\prime}\left(0\right)$ may be undefined, with different
right and left limits, but the odd symmetry forces $H_{odd}^{\prime}\left(0\right)$
to always be well-defined.

As the frequency heterogeneity of the pair increases, the pair of
fixed points shift progressively in the relative phase of the oscillators.
We refer to these states as near-synchronous and near-antiphase. For
small frequency heterogeneity $\omega_{1-2}>0$, the phase difference
in the near-synchronous state is approximately inversely proportional
to the slope, $\phi_{1-2}\approx\frac{\omega_{1-2}}{2KH_{odd}'\left(0\right)}$.
Note that this phase difference is in units of time, while to compare
phase-locked states across oscillators with different periods we should
evaluate phase in radians. To account for this we rescale both the
phase and the slope $H_{odd}^{\prime}$,
\[
\hat{\phi}_{1-2}=\phi_{1-2}\frac{2\pi}{T}\approx\frac{\omega_{1-2}}{2K\hat{H}_{odd}^{\prime}\left(0\right)},\ \mathrm{where\ }\hat{H}_{odd}^{\prime}\left(0\right)\equiv\frac{T}{2\pi}H_{odd}^{\prime}\left(0\right).
\]

For larger heterogeneity, the phase difference continues to increase
until $\omega_{1-2}$ is greater than the amplitude of $H_{odd}$
(red level in \figref{pair-example-Hodd}). At this point the fixed
points of (\ref{eq:pair}) are lost in a saddle-node bifurcation,
and the phase-locked state transitions to ``phase walk-through,''
with the phase of the cells slipping past each other (red curve in
\figref{pair-spikes} right). 

We can thus quantify the robustness of synchrony to heterogeneity
in two ways: the slope $\hat{H}_{odd}^{\prime}\left(0\right)$ gives
the strength of synchrony for small heterogeneity, while the amplitude
of $H_{odd}$ gives the critical heterogeneity at which the near-synchronous
state is lost. Below, we consider the robustness and stability of
near-synchronous phase-locking in more detail for the subthreshold
and spiking components of the resonate-and-fire interaction function.

\subsection{Subthreshold contribution}

For the analysis of synchrony in the resonate-and-fire model, we begin
by evaluating the contribution to the odd component of the subthreshold
interaction function $H_{sub}$, and return to the spiking component
in \subsecref{Spike-phase-interaction}. For the subthreshold interaction
function the odd component is typically dominated by the first Fourier
component $\hat{H}_{odd}\left(\hat{\phi}\right)\propto\sin\left(\hat{\phi}\right)$),
so its slope and amplitude, our two measures of robustness, are roughly
proportional. We first identify the condition for stable synchrony
(positive slope), and show that the subthreshold contribution is virtually
always synchronizing, with negative slope only possible along the
spiking regime boundary . We then explore the robustness of synchrony
as measured by the magnitude of the positive slope, identifying a
dependence on the shape of the voltage trajectory through the reset
voltage $v_{R}$.

We use the decomposition (\ref{eq:H comps}) of $H_{sub}$ into component
functions with approximate symmetry (see \figref{rf_Hcomps}) to simplify
our calculation of the slope. We consider the limit of small decay
($\lambda\ll1$) by expanding to first order in the decay parameter
(see calculation in \subsecref{Slope-of-interaction}).

\begin{equation}
\begin{alignedat}{1}\hat{C1}_{odd}^{\prime}(0) & =\lambda T-\cos T\sinh\left(\lambda T\right)\approx\lambda T\left(1-\cos T\right)>0,\\
\hat{C2}_{odd}^{\prime}(0) & =\left(\sinh(\lambda T)-\lambda\sin T\right)\approx\lambda\left(T-\sin T\right)>0,\\
\hat{S}_{odd}^{\prime}(0) & =T-\sin T\cosh\left(\lambda T\right)\approx T-\sin T>0.
\end{alignedat}
\label{eq:Hcomp slopes}
\end{equation}
The $O\left(1\right)$ contribution to $\hat{H}_{odd}^{\prime}\left(0\right)$
from $S$ is the primary subthreshold factor determining of the strength
and stability of synchrony. The sign of this slope is determined by
$A_{S}$, the coefficient of the $S$ component. Specifically, the
condition for stable synchrony is approximated to order $\lambda$
by
\[
A_{S}=\sin\left(\theta_{T}-\alpha\right)=\cos\left(\theta_{T}+\frac{\pi}{2}-\alpha\right)>0.
\]
Note that this condition differs only to order $\lambda$ from the
PRC singularity condition (\ref{eq:singularity}), 
\[
\cos\left(\theta_{H}-\alpha\right)=\cos\left(\theta_{T}+\frac{\pi}{2}+\tan^{-1}\lambda-\alpha\right)=0,
\]
which determines the location of both the grazing bifurcation for
hard reset and the saddle-node bifurcation for soft reset. Near these
boundaries, a negative slope can result from either $A_{S}<0$ or
from $A_{S}\approx0$ and $A_{C1}<0$ or $A_{C2}<0$. 

\begin{figure}
\subfloat[]{\begin{centering}
\includegraphics[height=0.3\paperheight]{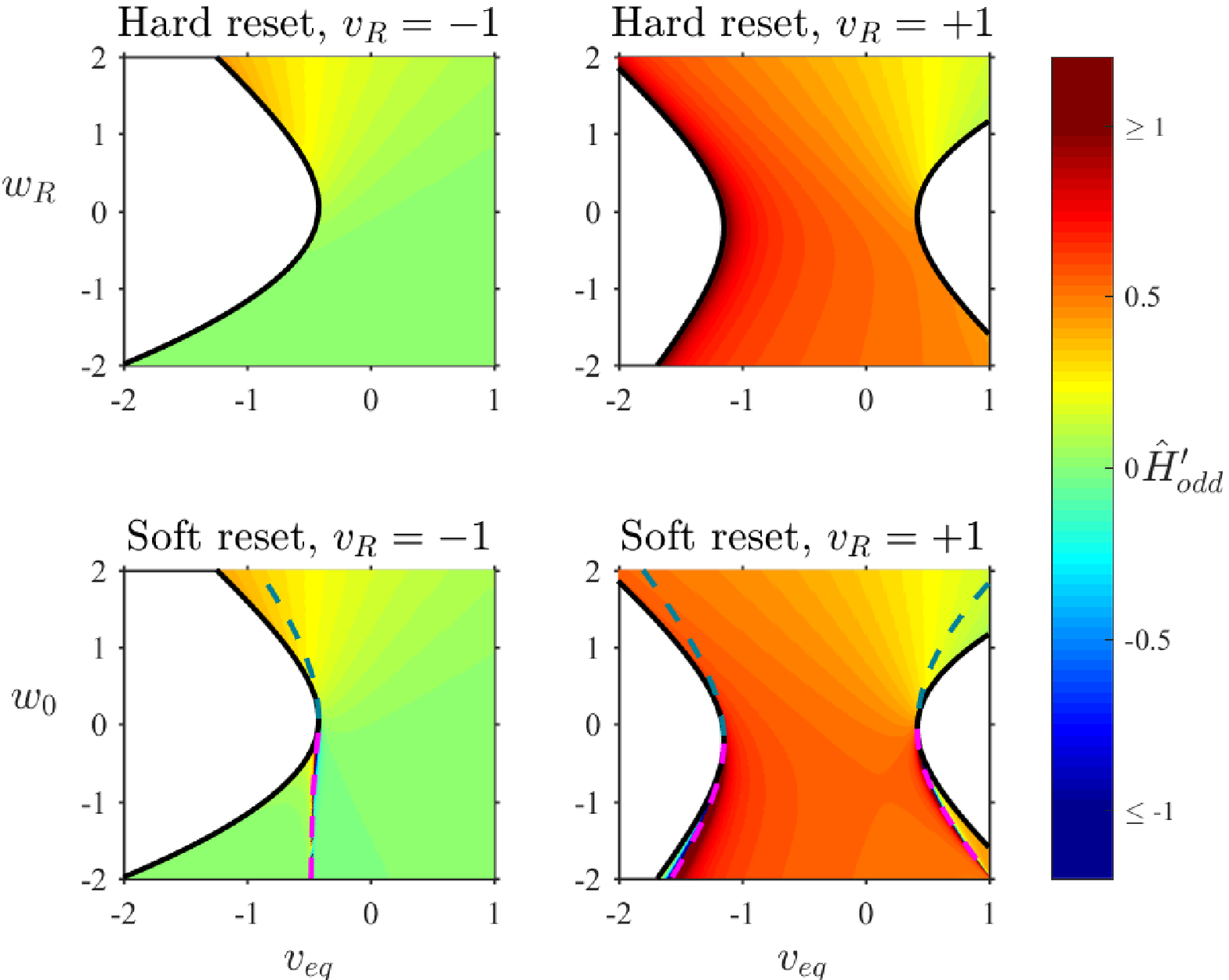}
\par\end{centering}
}\hfill{}\subfloat[\label{fig:Aodd_zoom}]{\begin{centering}
\includegraphics[height=0.3\paperheight]{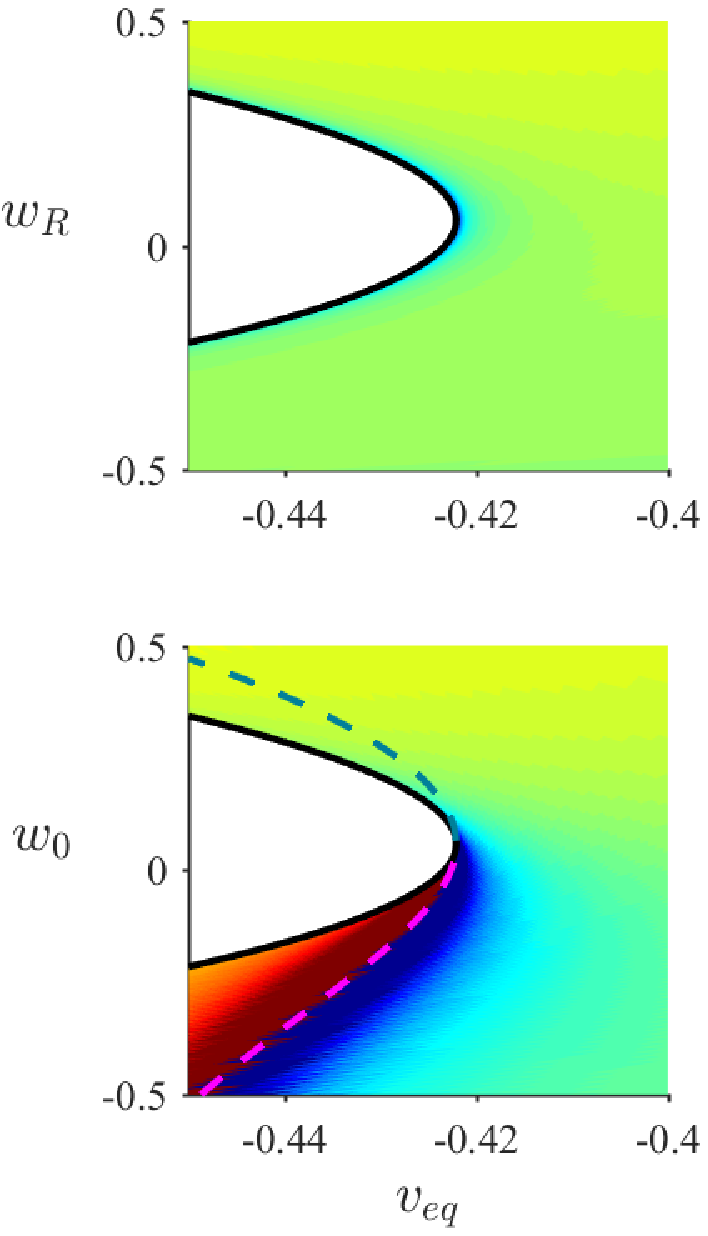}
\par\end{centering}
}

\caption{\label{fig:Aodd}Stability and robustness of synchrony. (a): Slope
of $H_{odd}$, the odd component of the subthreshold interaction function,
for $\lambda=0.1$. Magenta and cyan lines indicate stability boundaries
of the limit cycle. (b): Expansion of left column (negative reset
regime) near the spiking boundary.}
\end{figure}

We show the slope $\hat{H}_{odd}^{\prime}\left(0\right)$ for the
full parameter space in \figref{Aodd}, with the negative slope region
highlighted by zooming in near the spiking boundary (\figref{Aodd_zoom}).
Additionally, because the slope $\hat{H}_{odd}^{\prime}$ scales with
the diverging PRC amplitude $A$, both negative and positive slopes
near these boundaries can grow extremely large. We note, however,
that this result should be interpreted with caution, as the assumption
of weak coupling also breaks down approaching these boundaries.

\begin{figure}
\begin{centering}
\includegraphics[scale=0.6]{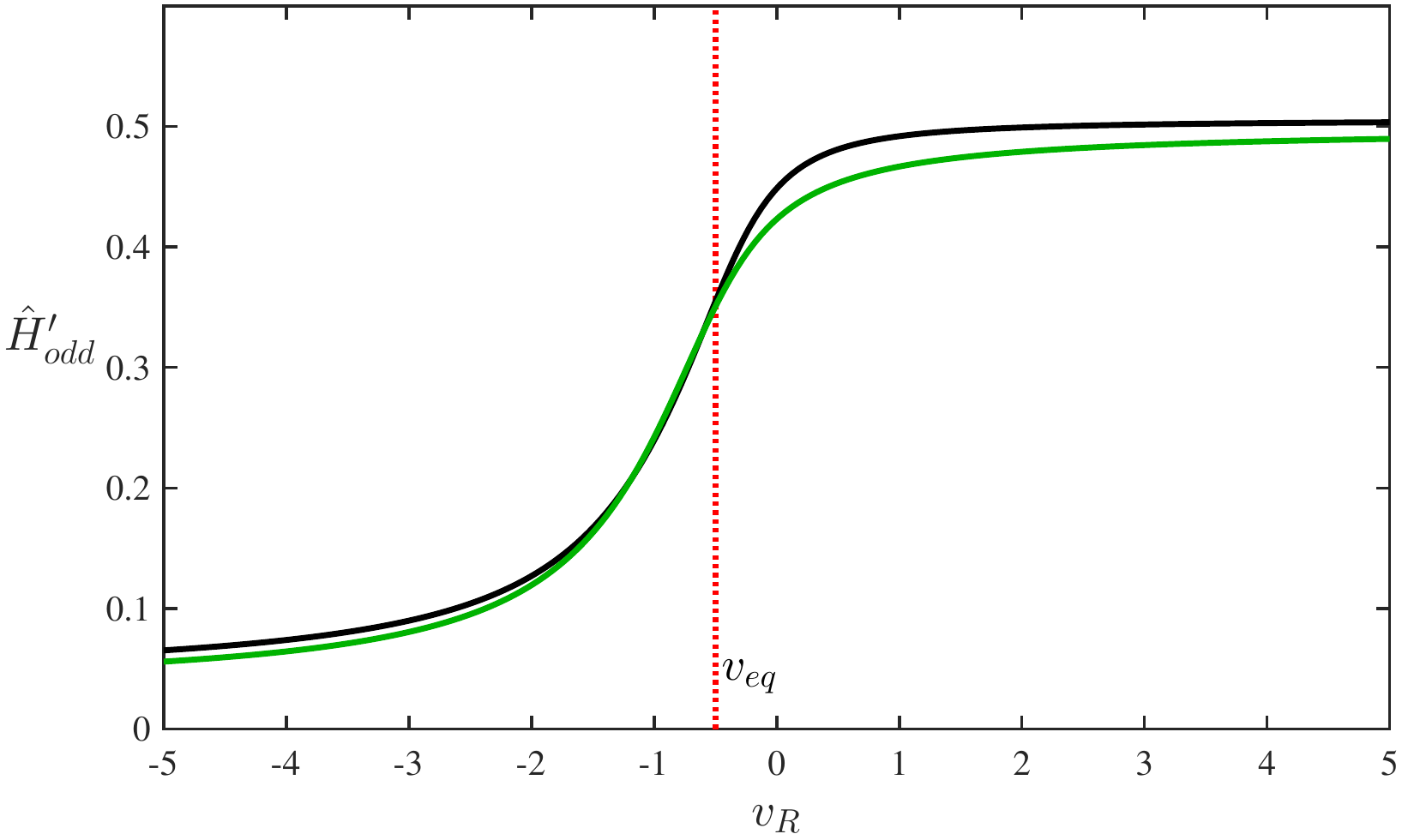}
\par\end{centering}
\caption{\label{fig:v0_trans}Robustness of synchrony (as measured by the slope
of $H_{odd}$) as the reset point $v_{R}$ is varied between the AHP
(negative extreme) and plateau potential (positive extreme) regimes.
Hard reset shown in black, soft reset in green. Parameters: $\lambda=0.1,\,w_{R}=1,\,v_{eq}=-0.5$.}
\end{figure}

The other significant trend in the slope, determining the robustness
of synchrony, is the difference between the positive and negative
reset regime. The slope is roughly unit magnitude in the positive
reset regime, sharply contrasting with the negative reset regime where
the slope is uniformly small (\figref{Aodd}). By explicitly plotting
the slope against the reset point $v_{R}$ in \figref{v0_trans},
we can see clearly the presence of two regimes with a distinct transition
in between. As shown in \figref{rf_traj} these two regimes have characteristic
voltage waveforms. A large positive $v_{R}$ is characterized by a
plateau potential in the voltage trajectory, while large negative
$v_{R}$ is characterized by after-hyperpolarization (AHP). We conclude
that for the resonate-and-fire model in the plateau potential regime,
the subthreshold dynamics significantly contribute to synchrony. In
the AHP regime, with little to no subthreshold contribution, the resonate-and-fire
model can only be synchronized by electrical coupling through the
effect of the transmitted spike (see \subsecref{Spike-phase-interaction}).

We can further investigate this trend through our slope approximation
(\ref{eq:Hcomp slopes}), focusing on the $S$ component as the dominant
slope contribution. As $v_{R}$ is increased, the coefficient $A_{S}$
varies little, while $\hat{S}_{odd}^{\prime}\approx T-\sin T$ increases
sharply with an increase in the period $T$. Note that with $\bar{\omega}=1$
fixed, $T$ represents two distinct factors: the extent of the cycle
in radians and its duration. However, varying the duration alone (via
$\omega$) has no effect on the phase-reduced dynamics (see \subsecref{app-Subthreshold-interaction-functio}),
so the effects on $H_{odd}$ are entirely due to the extent of the
limit cycle covering a larger fraction of a cycle of continuous oscillation.

\subsection{Spike interaction function\label{subsec:Spike-phase-interaction}}

The spiking component of the interaction function also has a significant
odd component, and thus contributes to the synchronization of the
coupled pair. We show here that this contribution always promotes
synchrony, reinforcing previous results for excitatory pulse coupling
of resonant neurons \citep{hansel1995SynchronyExcitatoryNeural}.

Because we model the spike as a $\delta$-function (with magnitude
$M$), the interaction function convolution equation (\ref{eq:phase_interaction})
has a simple form for the spiking component, i.e., a time-reversed
copy of the PRC.

\begin{align*}
H_{spike}\left(\phi\right) & =\frac{M}{T}\int_{0}^{T}Z_{v}\left(t\right)\left[\delta\left(t+\phi\right)-\delta\left(t\right)\right]dt,\\
 & =\frac{M}{T}\left(Z_{v}(T-\phi)-Z_{v}\left(0\right)\right)=\frac{MA}{Tr_{0}}e^{\lambda(T-\phi)}\cos\left(\phi-\alpha\right).
\end{align*}
The term $\delta\left(t\right)$ leads to an additional constant term
$Z_{v}\left(0\right)$, the value of the PRC at the spiking discontinuity.
Although the left and right limits of this point are defined by (\ref{eq:resfire prc}),
$Z_{v}\left(0\right)$ is unconstrained. In experiments and biophysically
detailed models, perturbations during a spike typically have little
to no effect, due to the many ion channels open during an action potential
lowering the input resistance of the cell \citep{gutkin2005PhaseResponseCurvesGive}.
Thus, we assume $Z_{v}\left(0\right)=0$, although we note that this
choice determines only a constant offset and does not affect the discontinuity
of $H_{spike}$.

Since the spike interaction function is discontinuous at zero, its
contribution to the stability and robustness of the synchronous state
depends primarily on the size and direction of the discontinuity.\footnote{If the amplitude of $H_{spike}$ is extremely large relative to $H_{sub}$,
it can create multiple local maxima of the interaction function, allowing
multiple stable states. In this case, our analysis still applies to
the synchronous phase-locked state, while other locked states would
require additional analysis.} The jump discontinuity can stabilize the fully synchronous state
even with nonzero heterogeneity, if $\left|\omega_{1-2}\right|<H_{odd}\left(0^{+}\right)=\Delta H_{spike}$
\citep{dodla2013EffectSharpJumps,lewis2012UnderstandingActivityElectrically}.
(For more in-depth analysis of the limit approaching discontinuity
in the interaction function, see Shirasaka et al. \citep{shirasaka2017Phasereductiontheory}.)
We evaluate this discontinuity directly from the PRC:

\begin{eqnarray*}
\Delta H_{spike} & = & H_{spike}\left(0^{+}\right)-H_{spike}\left(0^{-}\right),\\
 & = & \frac{M}{T}\left(Z(T^{-})-Z(0^{+})\right),\\
\Delta H_{spike} & = & \frac{MA}{Tr_{0}}\left(e^{\lambda T}\cos\alpha-\cos\left(T-\alpha\right)\right).
\end{eqnarray*}
The condition $\Delta H_{spike}>0$ is always satisfied for hard reset
($\alpha=0$), and in the soft reset case, $\Delta H_{spike}=0$ evaluates
to exactly the stability boundary (\ref{eq:cycle_instability}), with
$\Delta H_{spike}>0$ for all stable cycles. The spike interaction
function thus always promotes the stability of the synchronous state,
potentially synchronizing heterogeneous oscillators even when the
subthreshold contribution is not synchronizing. As we showed above,
the subthreshold contribution to the slope $H_{odd}^{\prime}\left(0\right)$
is near zero or negative for a significant portion of the parameter
space, including most of the AHP regime, thus requiring this spike
contribution in order to synchronize.

\section{Synchronization of a three-cell network: effect of the even component\label{sec:Even-3}}

Due to the symmetry of the coupled pair, the odd component of the
interaction function alone determines the evolution of the phase difference,
and the even-symmetric component has no effect on synchronization.
However, the even component has the potential to strongly affect synchronization
in both larger resonate-and-fire model networks and in actual biological
networks. Although many studies of synchronization in model neurons
ignore this possibility by focusing on symmetrically coupled pairs,
the effect of coupling with an even component has been explored in
large, sparse neuronal networks \citep{golomb2000NumberSynapticInputs}
and in large regular networks of phase oscillators (Kuramoto model
and generalizations) \citep{acebron2005Kuramotomodelsimple}. Here,
we show that similar effects can be seen in minimal networks, specifically
asymmetrically coupled pairs and three-cell networks, the study of
which can help us understand the more complex properties of larger
networks.

We first provide intuition on the effects of the even component by
revisiting the dynamics of a coupled pair, in terms of the phase difference
$\phi_{1-2}$ and the mean phase $\bar{\theta}$.
\begin{equation}
\begin{aligned}\frac{d\phi_{1-2}}{dt} & =\omega_{1-2}-2\bar{k}H_{odd}\left(\phi_{1-2}\right)+\Delta kH_{even}\left(\phi_{1-2}\right),\\
\frac{d\bar{\theta}}{dt} & =\bar{\omega}+2\bar{k}H_{even}\left(\phi_{1-2}\right).
\end{aligned}
\label{eq:pair_asymm}
\end{equation}
In the symmetrically coupled pair ($\Delta k=k_{12}-k_{21}$=0), the
even component has no effect on the phase difference, but it does
shift the frequency $\frac{d\bar{\theta}}{dt}$ of the phase-locked
state. In the case of asymmetric coupling strength $k_{12}\neq k_{21}$,
the term $\Delta kH_{even}\left(\phi\right)$ can be interpreted as
a differential shift in the instantaneous frequencies of the two oscillators.
This effective frequency shift can either promote or oppose synchrony
depending on whether the sign of the product adds to or cancels with
the intrinsic frequency heterogeneity $\omega_{1-2}$.

In a three-cell network, an even component term in the phase difference
dynamics can also arise from coupling to a third oscillator with a
different intrinsic frequency, even if the coupling is fully symmetric.
The phase difference equation for a symmetric three-cell network is
\[
\frac{d\phi_{i-j}}{dt}=\Delta\omega_{i-j}+K\left(\underset{-2H_{odd}}{\underbrace{H\left(-\phi_{i-j}\right)-H\left(\phi_{i-j}\right)}}+H\left(-\phi_{i-k}\right)-H\left(\phi_{i-k}-\phi_{i-j}\right)\right).
\]
While the first two $H$ terms partially cancel to isolate the odd
component, the latter two terms depend essentially on the even component.
We will first demonstrate the effects of these additional even component
terms on phase-locking, along with the effective frequency shifts
from coupling asymmetry, in the context of three-cell networks. We
will then return to the resonate-and-fire model, investigating the
size of the even component of the subthreshold interaction function,
its potential effects on neuronal networks, and its origin in the
model dynamics.

\subsection{Phase-locking of three cells\label{subsec:Phase-locking-of-three}}

A network of three cells is a minimal case that allows network asymmetry
(of coupling or frequency) to trigger an effect of the even component
even when the pairwise coupling is symmetric. In this context, the
amplitude of the odd component (relative to the heterogeneity) is
still the primary factor determining the existence of synchronous
phase-locking, but the addition of an even component can modify the
outcome dramatically, especially when the even component grows large
relative to the odd component.

The phase model for the three-cell network is assumed to take the
general form of (\ref{eq:phase_model}). We additionally assume pairwise
symmetry of the coupling, $k_{ij}=k_{ji}$, and expand the phase difference
equations for the network,

\begin{equation}
\begin{alignedat}{1}\dot{\phi}_{1-3} & =\Delta\omega_{1-3}+k_{21}H\left(-\phi_{1-2}\right)+k_{31}H\left(-\phi_{1-3}\right)-k_{31}H\left(\phi_{1-3}\right)-k_{23}H\left(\phi_{1-3}-\phi_{1-2}\right),\\
\dot{\phi}_{1-2} & =\Delta\omega_{1-2}+k_{31}H\left(-\phi_{1-3}\right)+k_{21}H\left(-\phi_{1-2}\right)-k_{21}H\left(\phi_{1-2}\right)-k_{23}H\left(\phi_{1-2}-\phi_{1-3}\right).
\end{alignedat}
\label{eq:triple phase diff}
\end{equation}
To simplify calculations in this analysis, we restrict the interaction
function to its first Fourier components,\footnote{For the resonate-and-fire subthreshold interaction function, higher
modes of the Fourier expansion contribute no more than 6\% of the
variance in the parameter spaces shown (in \figref{Aodd} and \figref{Aeven},
with $\lambda=0.1$).}
\begin{align}
\hat{H}\left(\hat{\phi}\right) & =A_{odd}\sin\hat{\phi}+A_{even}\left(1-\cos\hat{\phi}\right),\label{eq:H_firstfourier}
\end{align}
where $\hat{\phi}$ and $\hat{H}$ indicate phase in radians (note
that we drop the hat notation below). We also impose the constraint
that cosine be accompanied by a constant offset, $H_{even}\propto1-\cos\phi$,
from the diffusive coupling condition $H\left(0\right)=0$. We parametrize
the amplitude of the even component by fixing the odd component and
varying the amplitude ratio, 
\[
\beta=\tan^{-1}\frac{A_{even}}{A_{odd}}.
\]

\begin{figure}
\begin{centering}
\subfloat[\label{fig:even_small_nc}]{\hfill{}\includegraphics[scale=0.7]{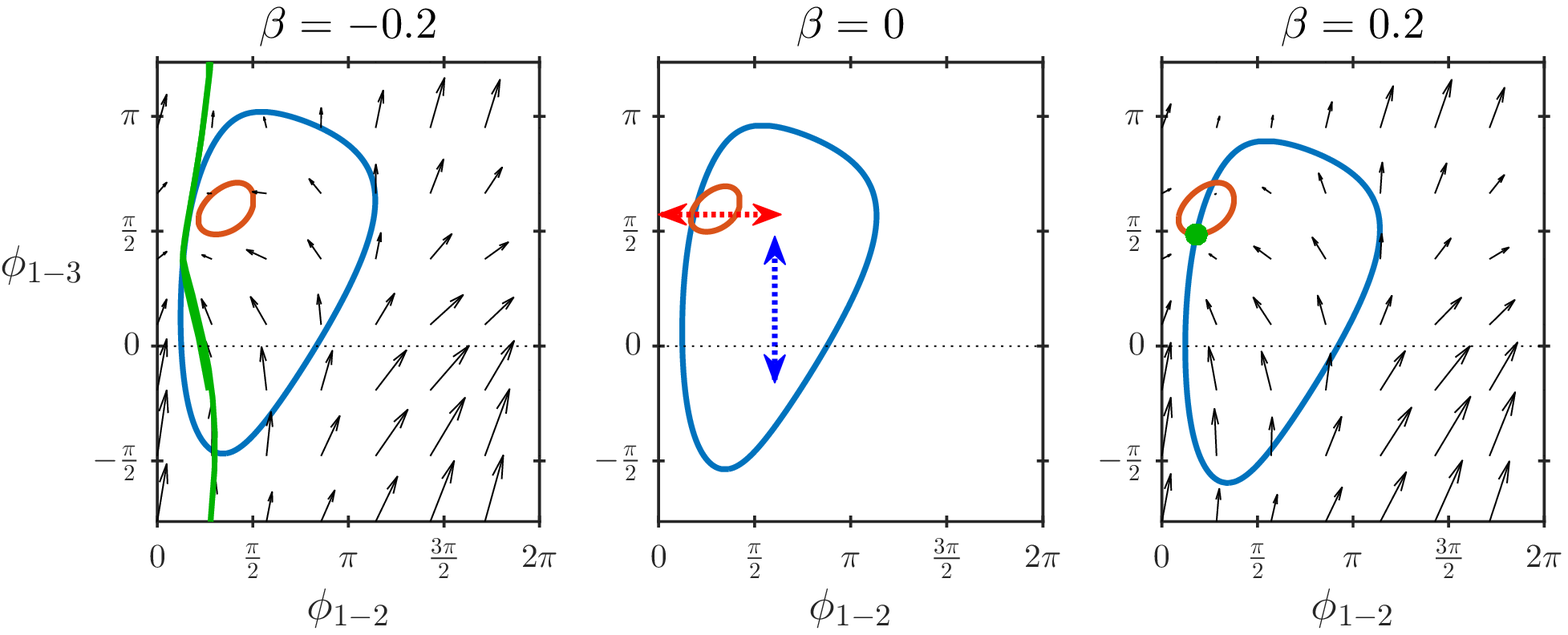}\hfill{}

}
\par\end{centering}
\begin{centering}
\subfloat[\label{fig:even_small_spikes}]{\includegraphics[scale=0.7]{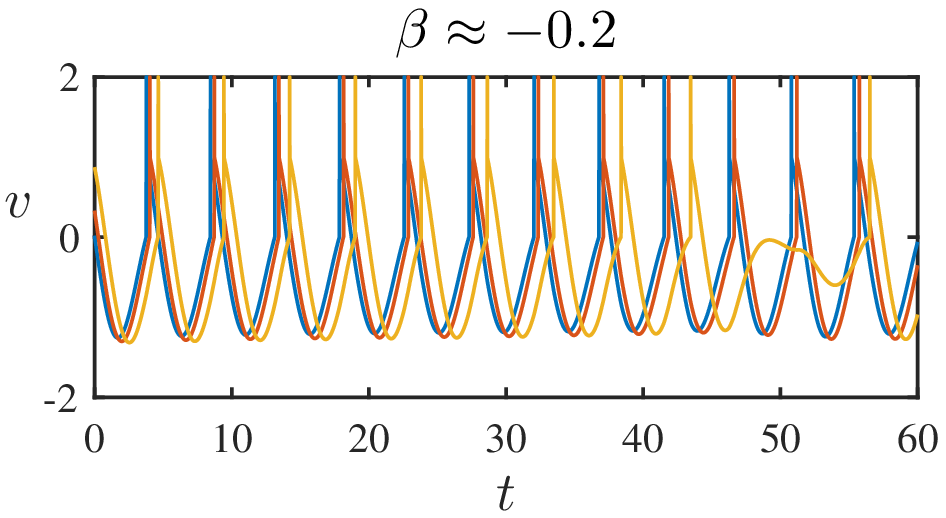}\hfill{}\includegraphics[scale=0.7]{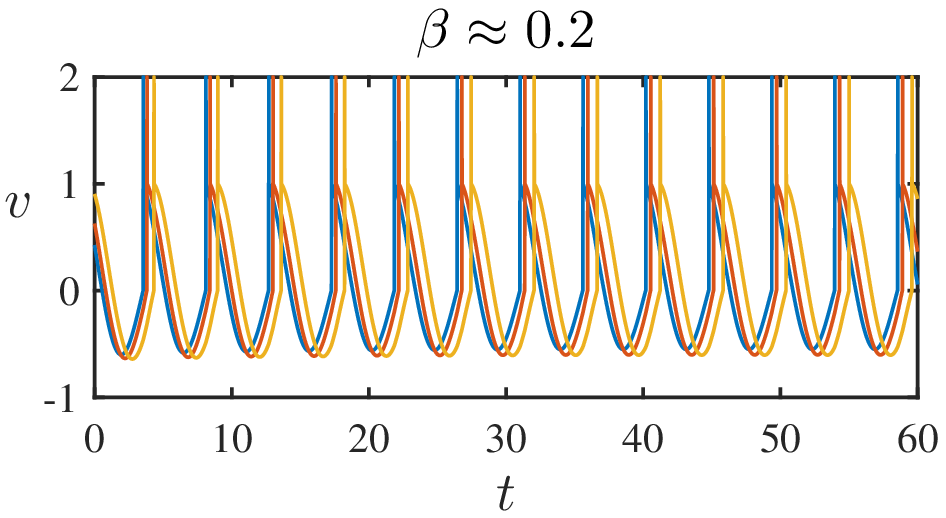}

}
\par\end{centering}
\caption{\label{fig:even_small}Effect of a small even component on phase-locking
of three-cell network. (a) Phase plane with ullclines of phase model
(blue for $\phi_{1-2}$, red for $\phi_{1-3}$), at the critical coupling
for existence of phase-locking with $\beta=0$. The shift of nullclines
eliminates the fixed point for negative $\beta$. Stable limit cycles
and fixed points shown in green. (b) Simulations of the resonate-and-fire
model show loss of fixed point with negative even component. Note
that the phase-slipping oscillator (yellow) also misses a spike. Blue,
red, and yellow denote oscillators 1, 2, and 3 respectively. Parameters:
$\omega=\left(1.067,\,1.017,\,0.917\right)$, $k_{ij}=0.09$ ($A_{odd}=0.044$),
$M=0$, $\lambda=0.1$, $v_{R}=1$. For $\beta$ positive/negative
respectively, $w_{R}=\left(0,\,0.49\right)$, $v_{eq}=\left(-0.03,\,-0.3\right)$.}
\end{figure}

Examples of the two-dimensional phase plane from equations (\ref{eq:triple phase diff})
are shown in \figref{even_small_nc}, for the phase reduction of the
resonate-and-fire model on a \emph{symmetric} three-cell network (approximated
in the form (\ref{eq:H_firstfourier})). The intersections of the
nullclines are fixed points of the system, corresponding to phase-locked
states. For $\beta=0$, the odd coupling strength $A_{odd}$ is set
at the critical value for phase locking of the network. Oscillators
1 and 2 are closely locked, while 1 and 3 are locked at $\phi_{1-3}\approx\frac{\pi}{2}.$
For small changes in the even component, $\beta>0$ shifts the nullclines
together to promote phase-locking, while $\beta<0$ shifts them apart.
Corresponding simulations of the full resonate-and-fire model network
are shown in \figref{even_small_spikes}. For $\beta<0$, oscillators
1 and 2 remain entrained, but oscillator 3 slips past in relative
phase. 

In a general three-cell network with arbitrary connection weights,
this shift of the nullclines combines with the synchronizing effect
of the odd component and with the effective heterogeneity from coupling
asymmetry to determine the presence or absence of phase locking. In
the limit of small $\beta$, we can show analytically that the shift
of the nullclines described above is the generic first-order effect
on symmetric networks, and separates additively from the frequency
and coupling heterogeneity terms when both are present. To do this,
we first simplify the form of the nullclines by rewriting the coupling
as a phase lag accompanied by an offset,
\begin{equation}
H\left(\phi\right)=A_{odd}\left(\sin\phi+\tan\beta\left(1-\cos\phi\right)\right)\approx A_{odd}\left(\sin\left(\phi-\beta\right)+\beta\right).\label{eq:H_beta}
\end{equation}
Taking one of the nullcline equations from (\ref{eq:triple phase diff}),
we then combine all the coupling terms into a single sine function,
capturing parametric dependence of the nullcline in the offset, amplitude,
and phase of this effective interaction. 
\begin{equation}
0=\frac{\omega_{1-3}}{A_{odd}}+\omega_{eff}+f_{C}+f_{A}\sin\left(\phi_{1-3}+f_{\alpha}\right),\label{eq:triple-eff}
\end{equation}
where 
\begin{align*}
\omega_{eff} & \approx\left(k_{21}-k_{23}\right)\beta,\\
f_{C} & \approx-k_{21}\sin\left(\phi_{1-2}+\beta\right),\\
f_{A} & \approx\sqrt{k_{23}^{2}+4k_{31}^{2}+4k_{23}k_{31}\cos\left(\phi_{1-2}+\beta\right)},\\
f_{\alpha} & \approx\arctan\left(\frac{-k_{23}\sin\left(\phi_{1-2}+\beta\right)}{2k_{31}+k_{23}\cos\left(\phi_{1-2}+\beta\right)}\right).
\end{align*}
The odd component affects only the term $\frac{\omega_{1-3}}{A_{odd}}$,
decreasing the effect of the intrinsic frequency heterogeneity. This
can be shown to increase the extent of the nullclines, promoting a
fixed point as with the coupled pair. The frequency shift term $\omega_{eff}$
gives the effective heterogeneity from coupling asymmetry. Similar
to the asymmetrically coupled pair (\ref{eq:pair_asymm}), this supports
or opposes phase locking depending on its sign relative to the intrinsic
heterogeneity $\omega_{1-3}$. 

\begin{figure}
\begin{centering}
\includegraphics[scale=0.7]{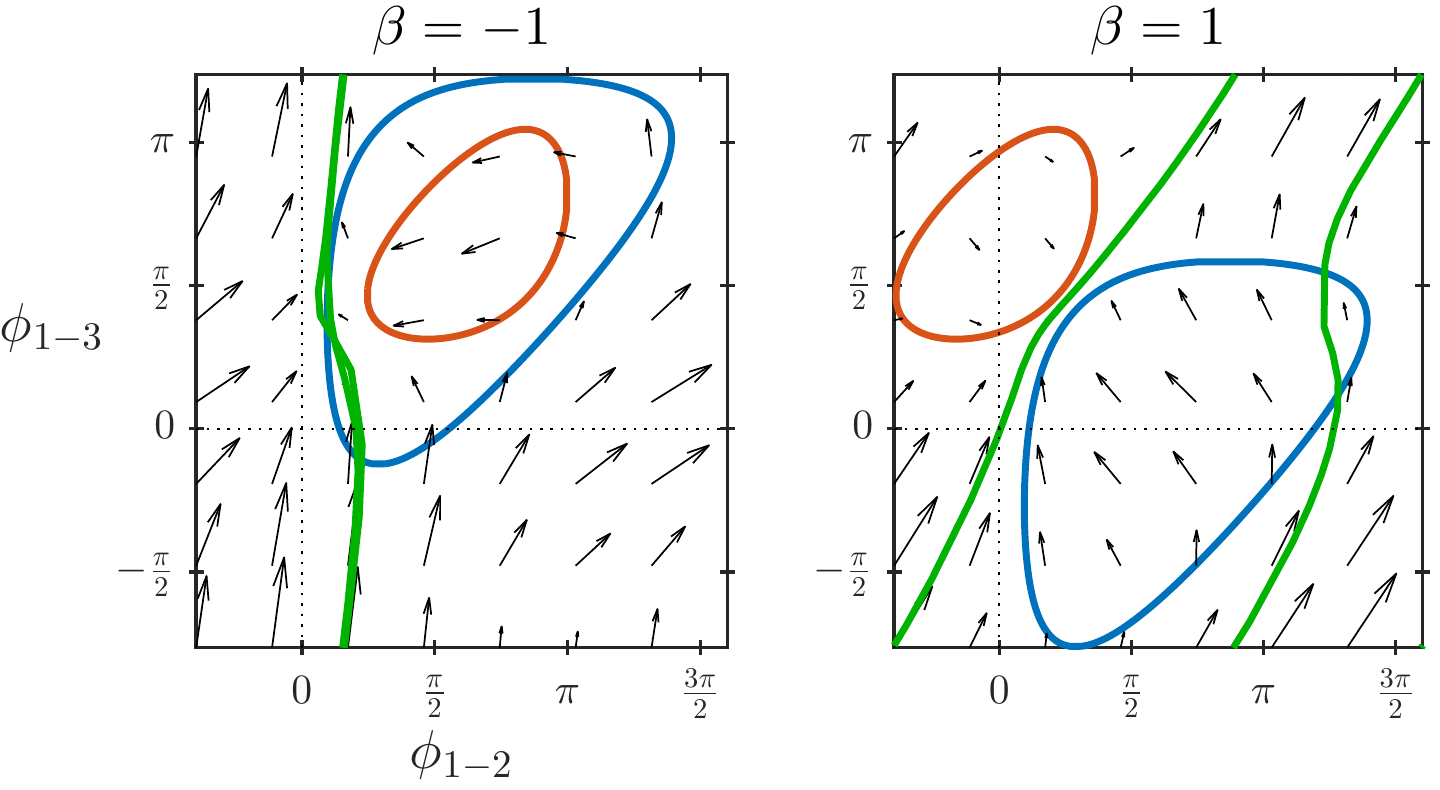}
\par\end{centering}
\caption{\label{fig:even_large}Effect of a large even component on phase-locking
of three-cell network. Nullclines of phase model (blue for $\phi_{1-2}$,
red for $\phi_{1-3}$), at the critical coupling for existence of
phase-locking with $\beta=0$, are dramatically shifted and distorted
for both large positive or negative even component, eliminating the
fixed point in both cases. Stable limit cycles and fixed points shown
in green: for $\beta=-1$ oscillators 1 and 2 are entrained, and for
$\beta=1$ oscillators 2 and 3 are entrained at a 1:2 frequency ratio.
Parameters as in \figref{even_small}.}
\end{figure}

The remaining terms represent the additional effects of the even component
through their joint dependence on $\beta$ and $\phi_{1-2}$, which
for small $\beta$ reduces to a dependence on the sum $\left(\phi_{1-2}+\beta\right)$.
In a network with symmetric coupling ($\omega_{eff}=0$), this dependence
on $\left(\phi_{1-2}+\beta\right)$ causes the $\phi_{1-3}$ nullcline
to shift along the $\phi_{1-2}$ axis with a change in $\beta$ (and
likewise for the $\phi_{1-2}$ nullcline with respect to $\phi_{1-3}$).
In the absence of frequency heterogeneity, the nullclines are straight
lines $\phi_{1-3}=0$ and $\phi_{1-2}=0$ (see (\ref{eq:triple phase diff})),
and thus are unaffected by these shifts. For larger frequency heterogeneity,
the nullclines form closed curves and a phase-locked fixed point can
be lost in a saddle-node bifurcation if the shifts move the nullclines
apart. This can occur near the bifurcation for a small change in $\beta$
of the proper sign (relative to the frequency heterogeneity) as shown
in \figref{even_small}, or for larger changes in $\beta$ regardless
of sign. As $\beta$ grows larger, the shift of the nullclines increases
proportionally (accompanied by distortion from higher order terms
not included in \ref{eq:triple-eff}). An example where this larger
shift eliminates the synchronous fixed point regardless of the sign
of $\beta$ is shown in \figref{even_large}. 

\begin{figure}
\begin{centering}
\includegraphics[scale=0.6]{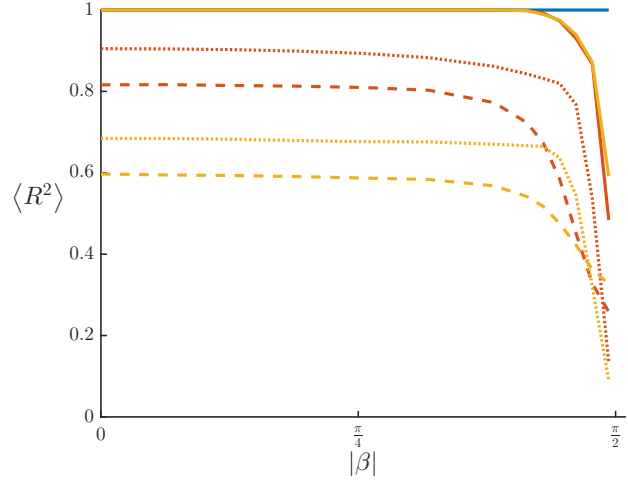}
\par\end{centering}
\caption{\label{fig:even-effect-avg}Average level of synchrony $R^{2}$ for
random three-cell networks decreases with increasing amplitude of
even component. Note that $\text{\ensuremath{\left|\beta\right|}=\ensuremath{\frac{\pi}{2}} }$
is the limit of infinite even coupling. Dashed lines indicate both
frequency and coupling heterogeneity; solid lines, coupling only;
dotted lines, frequency only. Colors show level of heterogeneity.
Frequencies are drawn from a Gaussian distribution (mean 0) and coupling
weights from a log-normal distribution (mean 1), both with standard
deviation $\sigma=0,\,1,\,2$ for blue, red, yellow respectively..
Odd component of coupling fixed at $A_{odd}=1$.}
\end{figure}

To generalize from these minimal network examples to dynamics on larger
networks, we investigate which effects of the even component occur
generically for random networks versus contingent on the specific
coupling and frequencies. The global synchrony of a large network
will in some sense average the synchrony of random subnetworks, so
we explicitly average over random three-cell networks. In \figref{even-effect-avg}
we plot the Kuramoto order parameter $R^{2}=\frac{1}{N_{t}}\sum_{t}\left|\frac{1}{N_{j}}\sum_{j}e^{i\phi_{j}\left(t\right)}\right|^{2}$,
a measure of the strength of synchrony, averaged over time and over
instantiations of random frequencies and coupling heterogeneity. Effects
that depend on the sign of $\beta$ (relative to the frequency or
coupling heterogeneity) are averaged out; we see no effect for small
$\beta$ of either the nullcline shifts or coupling asymmetry effects.
For large $\beta$, we see a significant decrease in synchrony with
both types of heterogeneity, as the even component effects begin to
dominate over the intrinsic frequency heterogeneity (as in \figref{even_large}).
This result resembles analytical results for chains of oscillators
with spatially constrained coupling, where an increase in the critical
coupling occurs for $\beta\geq\frac{\pi}{4}$ \citep{omelchenko2014Partiallycoherenttwisted,wolfrum2016TurbulenceOttAntonsen}.
It should be noted that in large networks the effect of the even component
on synchronization can depend dramatically on the probability distribution
from which the frequencies are drawn \citep{omelchenko2012NonuniversalTransitionsSynchrony},
whereas this dependence for random three-cell networks is minimal.

\subsection{Even component of subthreshold interaction function}

\begin{figure}
\begin{centering}
\includegraphics[scale=0.8]{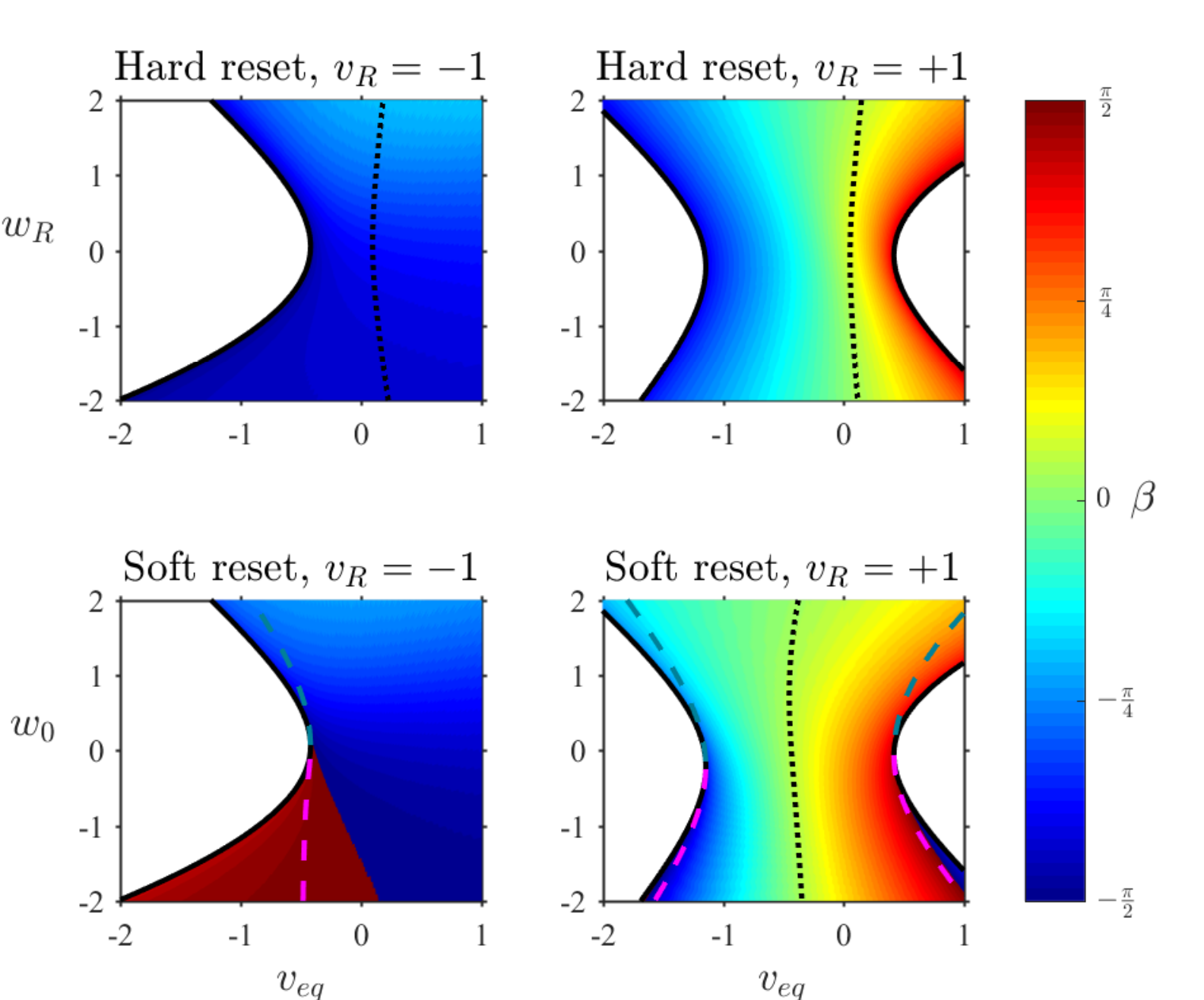}
\par\end{centering}
\caption{\label{fig:Aeven}Relative amplitude $\beta$ of even to odd components
of the resonate-and-fire subthreshold interaction function, from least-squares
fit of $H$ to the form (\ref{eq:H_beta}) (for $\lambda=0.1$). Magenta
and blue-green dashed lines indicate stability boundaries of the limit
cycle for positive and negative slope instabilities. Dotted line indicates
condition (\ref{eq:no shear}) for no reset-induced shear. }
\end{figure}

In the previous section we showed that the even component of the interaction
function can have significant effects on phase locking. We now proceed
to assess the magnitude of the even component for the resonate-and-fire
oscillator. In \figref{Aeven}, we plot the amplitude ratio $\beta$
of the even component for the subthreshold interaction function. We
note a dramatic difference between the positive and negative reset
regimes (as seen with the odd component slope in \figref{Aodd}).
$\beta$ is consistently large and negative in the negative reset
regime (except for a small positive region for soft reset where the
odd component is negative), and varies significantly across the positive
reset regime. Although both the odd and even amplitudes decrease with
$v_{R}$ (towards $v_{R}=-1$), the odd component remains smaller,
explaining the large amplitude ration in the negative reset regime.
However, since the overall amplitude of the subthreshold interaction
function is small, any effects of the even component here are likely
to be dominated by the spike interaction, as discussed in \subsecref{Spike-phase-interaction}.
In the positive reset regime, both the overall and relative amplitudes
can be significant. $\beta$ increases from negative to positive with
increases in $v_{eq}$, and is largest in magnitude at the spiking
regime boundaries. This trend is driven by the component function
$C1$ from (\ref{eq:H comps}), for which the coefficient $A_{C1}\approx\frac{A}{2}v_{eq}$
(for hard reset). The component function $C2$ also contributes to
the even component, but to a lesser degree.

Where the even component is small in magnitude (\figref{Aeven}, pale
color region) the effects depend on the sign of $\beta$ and cancel
when averaged over random networks. This is unlikely to significantly
affect large biological networks. For local adaptation to overcome
this averaging and\emph{ }support or oppose synchrony, the even component
of a connection would need to adapt based on frequencies and coupling
strengths of both coupled cells, a possibility that seems biologically
unrealistic. However, the even component is sufficiently large in
a significant portion of the parameter space to potentially oppose
synchrony in random networks (as in \figref{even-effect-avg}). In
the neural context, for systems in which synchrony supports biological
function, cells may need to adapt their dynamical properties to keep
the even component small. This can occur most directly through shifting
the equilibrium $v_{eq}$, either by slower changes in conductances
or synaptic weights, or by faster shifts in the tonic input to the
cell. This mechanism could potentially enable rapid adaptive shifts
(up or down) in the level of synchrony. Finally, it is possible that
a large even component could instead promote specific functional states,
such as a chimera state \citep{laing2009dynamicschimerastates}, rather
than simply opposing synchrony.

\subsection{Origin of the even component}

In some of the mathematical literature on coupled oscillators, strictly
odd-symmetric coupling is assumed as an easier to analyze default
case {[}e.g., \citealp{dorfler2014Synchronizationcomplexnetworks,ermentrout1984FrequencyPlateausChain}{]}.
Our analysis shows that, for the resonate-and-fire model, strictly
odd coupling is certainly not generic. To understand the factors determining
the variation in the even component, it helps to consider the general
case as arising from phase shifts relative to the purely odd special
case. Consider an approximation in which the limit cycle, PRC, and
resulting interaction function are all sinusoidal, differing only
in phase shifts. (This is strictly true only in the small decay and
large period limit.) The phase of the interaction function follows
from the relative phase of $v$ and $Z_{v}$, determined by the boundary
condition (\ref{eq:boundary}) for the adjoint equation. For the hard
reset, if the limit cycle crosses the threshold at $\theta_{T}\approx\nicefrac{3\pi}{2}$
($v_{eq}\approx0$), $v$ and $Z_{v}$ are out of phase by $\approx\frac{\pi}{2}$,
resulting in an odd interaction function $H\approx\sin$. Any phase
shift in the PRC away from this produces a corresponding shift of
the interaction function, introducing a nonzero even component. This
can occur if $v_{eq}$ shifts away from 0, or from the soft reset
boundary condition phase shift $\alpha$. 

A more geometrical perspective on the relative phase of the limit
cycle and PRC is through the concept of dynamical \emph{shear}, variation
of angular velocity with radial displacement from a limit cycle. In
a model without shear, perturbations perpendicular to the cycle do
not cause phase shifts, so the limit cycle is normal to its isochrons
and parallel to the (vector) PRC. This is close to the condition that
components of the limit cycle and PRC be $\frac{\pi}{2}$ out of phase,
and tends to lead to an odd interaction function for diffusive coupling.
For instance, in the Stuart-Landau oscillator, a minimal model for
shear about a limit cycle, the shear term in the dynamics directly
scales a cosine term in the interaction function \citep{aronson1990Amplituderesponsecoupled}.
In the resonate-and-fire model, although there is no shear in the
linear subthreshold dynamics, the effect of the threshold shifts the
orientation of the isochrons relative to the limit cycle exactly like
shear in the Stuart-Landau oscillator. This effective ``reset-induced
shear'' is caused by perturbed trajectories on one side of the limit
cycle crossing threshold earlier than those on the other side. In
the soft reset case, this effect also depends on the geometry of the
reset manifold. We can assess the validity of this explanation by
evaluating a condition for no shear in the resonate-and-fire model,
\begin{equation}
Z\left(T\right)\propto\frac{dx}{dt}\left(T\right).\label{eq:no shear}
\end{equation}
In \figref{Aeven}, we show condition (\ref{eq:no shear}) for zero
reset-induced shear as a dotted line. We see this intuition breaks
down in the negative reset regime, where the odd component is extremely
small. Otherwise, the condition closely approximates $\beta=0$, with
small variations that result from effects of the discontinuities in
$v$ and $Z_{v}$ not accounted for in this analysis.

\section{Discussion}

We applied the theory of weakly coupled oscillators to study the synchronization
of resonate-and-fire neurons coupled by electrical synapses. The use
of a minimal hybrid model to capture the resonant dynamics allowed
much of this analysis to proceed analytically. We calculated the phase
reduction of the resonate-and-fire model neuron using the adjoint
method for the PRC of hybrid models, following Shirasaka et al. \citep{shirasaka2017Phasereductiontheory}.
We also presented a simplified derivation of their technique. We found
that the resonant properties give rise to a potentially strong contribution
of the subthreshold fluctuations to synchronization, in addition to
the synchronizing effect of the spike. We also showed that, despite
having no effect on coupled pairs, effects arising from the reset
(i.e. reset-induced shear) have the potential to impair synchronization
in certain network configurations.

\subsection{Synchronization of Resonate-and-fire oscillators}

Our analysis focused on the resonate-and-fire oscillator as an idealized
model to study synchronization of electrically coupled resonant neurons.
The phase reduction technique allowed us to separate the interaction
function into components from spiking and subthreshold voltage fluctuations,
dissecting their distinct contributions to synchronization. We showed
that, in the resonate-and-fire model, the effect of a fast voltage
spike transmitted through electrical coupling is always synchronizing,
reinforcing previous work on pulse coupling of resonant neurons \citep{hansel1995SynchronyExcitatoryNeural,ladenbauer2012ImpactAdaptationCurrents,ermentrout2001EffectsSpikeFrequency,stiefel2009effectscholinergicneuromodulation}.
The contribution from subthreshold fluctuations generally promotes
synchrony as well, but can actively oppose synchrony in a small region
of parameter space. Where the spike and subthreshold synchronizing
effects oppose one another, our model predicts that the net effect
will depend linearly on the relative magnitude of the subthreshold
fluctuations and the spike. Although our model does not explore the
factors determining spike size, this can be inferred from more detailed
biophysical models and applied to a more quantitative analysis of
the resonate-and-fire spike effects.

In coupled pairs and other networks with extensive symmetry, synchronization
is solely determined by the odd component of the interaction function.
For the resonate-and-fire model, the subthreshold contribution to
this odd component generally has a positive slope, promoting stable
synchrony. This subthreshold contribution is small when the reset
voltage $v_{R}$ is strongly negative (corresponding to after-hyperpolarization).
A stronger contribution to the odd component supporting synchronization
occurs when $v_{R}$ is well above threshold (corresponding to a plateau
potential). The only significant departures from this rule are strong
subthreshold effects near the boundaries of the spiking regime, including
the small region with desynchronizing effects. However, the assumption
of weak coupling breaks down near these bifurcations, so this conclusion
should be verified by different methods of analysis.

Finally, we showed that in networks with less symmetry, significant
reset-induced effects on synchronization can appear. The even component
of the interaction function is often ignored, both because analyses
focus on symmetrically coupled pairs (e.g., \citep{lewis2012UnderstandingActivityElectrically})
and because, as observed by Sakaguchi \citep{sakaguchi1988Mutualentrainmentoscillator},
it has complex ``ambivalent effects on mutual entrainment.'' We
analyze three-cell networks to show how the even component has the
potential to oppose synchrony, especially when large enough to dominate
over intrinsic frequency heterogeneity. In the resonate-and-fire model,
the even component varies strongly with the equilibrium voltage, potentially
interfering with the subthreshold synchronizing effect in parts of
the positive reset regime. In general, any phase shift of the interaction
function will introduce an even component; our derivation of the adjoint
method for hybrid model PRCs clarifies a mechanism for such phase
shifts linked to the post-spike reset. The boundary condition for
the hybrid PRC determines the phase shift, dependent on the reset
map (hard or soft reset) and on the geometry of the trajectory, threshold,
and reset manifold. We characterize this effect as ``reset-induced
shear'': a phase shift results when trajectories on one side the
limit cycle cross threshold and are reset ahead of the limit cycle
trajectory. 

\subsection{Comparison of resonator and integrator neurons}

Taken as a whole, our results show that subthreshold resonance of
model neurons can have a significant synchronizing effect in electrically
coupled networks, contrasting with typical observations of integrator
neurons. Previous work on single-variable integrate-and-fire models
has found that the subthreshold effect of electrical coupling tends
to oppose synchrony \citep{lewis2003Dynamicsspikingneurons,pfeuty2003ElectricalSynapsesSynchrony}.
Because the reset voltage must be below threshold, the effect of the
reset is desynchronizing \citep{lewis2012UnderstandingActivityElectrically}
and tends to dominate the small synchronizing effects of other subthreshold
fluctuations. Thus, in simple integrator models synchronization must
rely on transmission of the spike only, functionally similar to pulse
coupling from fast excitatory chemical synapses. In contrast, electrically
coupled resonator neurons may combine pulse and continuous coupling
to synchronize. This may help explain experimental observations of
a loose correlation across brain regions between resonant properties
of neurons and the prevalence of electrical synapses \citep{hutcheon2000Resonanceoscillationintrinsic,pereda2013Gapjunctionmediatedelectrical}.

We found that the subthreshold contribution to synchrony is strongest
in the plateau potential regime. Since the PRC is not dramatically
different between the plateau and AHP regimes, our analysis suggests
that the subthreshold synchronizing effect of resonance is mediated
primarily by the temporal extent of the voltage fluctuations. In the
plateau regime the subthreshold voltage waveform extends close to
a full sinusoidal cycle, providing greater opportunity for exchange
of current. This synchronizing effect likely extends beyond our resonate-and-fire
analysis to the electrical coupling of other resonator neurons. Experimental
results show plateau potentials in resonant neurons with widespread
electrical coupling in the inferior olive \citep{llinas1981Electrophysiologymammalianinferior,marshall2004InferiorOliveOscillations},
suggesting a potential synchronizing effect of the plateau. Synchronization
of subthreshold oscillations in the absence of spiking \citep{leznik2002Electrotonicallymediatedoscillatory}
may also rely on a similar mechanism. Our predictions concerning resonance
and subthreshold effects are directly testable experimentally, using
pharmacological manipulation of resonant properties or dynamic clamp
techniques to perturb and test single neurons and circuits, supplemented
by analysis of detailed biophysical models.

We note, however, that integrator versus resonator is not a strict
classification and does not always correspond directly with synchronization
properties, despite the general trends observed. Although type I excitability
(associated with a SNIC bifurcation), type I PRCs ($Z_{v}$ strictly
positive), and the integration of input are often taken as loosely
equivalent properties, Ermentrout et al. \citep{ermentrout2012shapephaseresettingcurves}
clarified that systems near a SNIC bifurcation can have type II PRCs,
with strong negative lobes. Additionally, Dodla and Wilson \citep{dodla2013EffectSharpJumps},
analyzing synchrony based only on the shapes of the PRC and voltage
fluctuations, emphasize that the type of PRC alone is insufficient
to determine the synchronization of electrically coupled oscillators.
Our work reinforces these results, showing that a resonator model
can in certain regimes have integrator-like properties, both in the
PRC and in the interaction function and synchrony. 

\subsection{Synchronization of hybrid model neurons}

Our use of an idealized hybrid model neuron for this study necessitated
modification of the PRC analysis techniques typically applied to biophysically
detailed continuous models (\subsecref{Hybrid-PRC}). The change in
perspective from continuous to discontinuous dynamics may help provide
new insights into basic questions of synchronization, such as whether
resonance or other properties of neurons support the synchronizing
effects of the spike. The discontinuous hybrid model PRC leads to
a spike interaction function that is discontinuous at the origin,
creating an especially strong synchronizing effect when the PRC jump
is negative (positive for the interaction function). On the other
hand, estimates of the (infinitesimal) PRCs from real neurons or biophysical
models are continuous and approximately zero at the instant of spiking.
If we smooth a hybrid neuron PRC to match these observations, the
negative jump translates to a smooth peak skewed ``rightward,''
to the latter portion of the PRC closely preceding the spike \citep{lewis2012UnderstandingActivityElectrically}.
Realistic PRCs generally show this rightward skew, which gives a synchronizing
positive slope to the interaction function, matching the strictly
positive resonate-and-fire discontinuity. The skew has been shown
to vary with adaptation in a range of models and experiments \citep{ermentrout2001EffectsSpikeFrequency},
including in hybrid models \citep{pfeuty2003ElectricalSynapsesSynchrony,ladenbauer2012ImpactAdaptationCurrents}.
Future work could further link these adaptation skew effects and the
resonance effects that we study, as well as bridging the gap between
hybrid and continuous models by explicitly considering specific spike
shapes along with the hybrid model dynamics.

For the subthreshold effects of electrical coupling, the interaction
function depends on both the PRC and the limit cycle. This allows
for significant variation in both odd and even components, even for
the simple resonate-and-fire model. In other hybrid models, the odd
component subthreshold effects have been shown to vary widely between
models and with variation of parameters \citep{pfeuty2003ElectricalSynapsesSynchrony,coombes2009GapJunctionsEmergent},
consistent with the diversity of electrical coupling effects on synchrony
(from combined spike and subthreshold effects) in many biophysical
models \citep{kepler1990effectelectricalcoupling,chow2000DynamicsSpikingNeurons,mancilla2007SynchronizationElectricallyCoupled}.
Our work reinforces this observation for the odd component and also
emphasizes similar variability in the even component, which likely
generalizes to other hybrid models. Specifically, reset-induced shear
is a newly identified factor that introduces a variable even component
to the interaction function. Our analysis techniques allowed us to
link this to effects of the reset map and the geometry of threshold
and reset, and can be applied more generally to disentangle complex
odd and even component effects in other hybrid models.

Despite the many advantages of hybrid models, little is formally known
about the bounds on their validity. Ideally one should have a rigorous
understanding of the hybrid model as a suitable asymptotic limit of
related continuous models, tying the reset map to the detailed dynamics
of spiking \citep{jolivet2004GeneralizedIntegrateandFireModels}.
Some work has touched on this for simple cases: comparing input response
\citep{smith2000FourierAnalysisSinusoidally,brunel2003Firingrateresonancegeneralized},
spiking transitions \citep{engelbrecht2009Dynamicalphasetransitions},
or network spiking dynamics \citep{hansel1998numericalsimulationsintegrateandfire}
between hybrid and biophysical models. Still, any direct comparison
between idealized hybrid model dynamics and more complex biophysical
models is challenging, whether at the single-cell or population level.
Phase reduction provides a possible locus for such comparison, since
detailed models can be phase-reduced computationally, translating
them into the same ``language'' as our study of the resonate-and-fire
model. The geometric insight into the PRC from our derivation of the
adjoint method for hybrid models facilitates the interpretation of
such comparisons. Our conclusions regarding the synchronization of
resonant neurons can thus be verified and extended by comparisons
with the computational phase reduction of detailed biophysical models
and with the empirical phase response analysis of real neurons.

\appendix

\section{~}

\subsection{\label{subsec:Connection-to-shira}Connection to Shirasaka et al.
\citep{shirasaka2017Phasereductiontheory}}

Here we will demonstrate that the boundary condition (\ref{eq:boundary})
for the hybrid model PRC across the reset discontinuity, which we
derived in \subsecref{Hybrid-PRC}, matches the result derived by
Shirasaka et al. \citep{shirasaka2017Phasereductiontheory} following
techniques from nonsmooth dynamical systems theory. The primary difference
between the two results is that we present $N-1$ boundary conditions,
for a reset map $R$ defined on the $\left(N-1\right)$-dimensional
threshold manifold, while Shirasaka et al. defined a reset map $\Phi$
in $N$ dimensions (on an open neighborhood of the threshold) and
presented $N$ distinct boundary conditions for the adjoint problem.
We show here that the $N-1$ conditions corresponding to our result
match exactly, and that the remaining condition simply enforces the
normalization (\ref{eq:normalize}) (regardless of the definition
of $\Phi$ off the threshold manifold). 

Their result is formulated in terms of the \emph{saltation matrix}
$C$,
\begin{equation}
Z\left(T^{-}\right)=C^{T}Z\left(T^{+}\right),\label{eq:saltation_boundary}
\end{equation}
\begin{equation}
C=D\Phi-M.
\end{equation}
\[
M=\left(D\Phi f\left(T^{-}\right)-f\left(T^{+}\right)\right)\frac{\hat{v}^{T}}{f_{v}\left(T^{-}\right)},
\]
where $f_{v}$ is the $v$-component of the dynamics, $\hat{v}$ is
a $v$-direction unit vector, and $T^{-}$ and $T^{+}$ are the left
and right limits of the boundary crossing. The Jacobian $D\Phi$ corresponds
to our directional derivatives $D_{u}R$ along threshold. The row
space of matrix $M$ is the $\hat{v}$ direction only (perpendicular
to the threshold), so for any component $Z_{i}$ along the threshold,
$M$ does not contribute to the boundary condition and (\ref{eq:saltation_boundary})
reduces to our result (\ref{eq:boundary}).
\begin{equation}
Z_{i}\left(T^{-}\right)=D_{i}\Phi^{T}Z\left(T^{+}\right)=D_{i}R^{T}Z\left(T^{+}\right).\label{eq:salt_bc_match}
\end{equation}

The remaining $v$-component boundary condition from (\ref{eq:saltation_boundary})
can be shown to simply enforce the normalization condition (\ref{eq:normalize}).
We first evaluate this final component,

\begin{align*}
Z_{v}\left(T^{-}\right) & =\frac{1}{f_{v}\left(T^{-}\right)}\Bigl(f_{v}\left(T^{-}\right)D_{v}\Phi-D\Phi\cdot f\left(T^{-}\right)+f\left(T^{+}\right)\Bigr)\cdot Z\left(T^{+}\right).
\end{align*}
We then expand $\left(D\Phi f\left(T\right)\right)\cdot Z\left(T^{+}\right)$
as a sum over components $\Sigma_{i=1}^{N}f_{i}\left(T^{-}\right)D_{i}\Phi\cdot Z\left(T^{+}\right)$.
If the $N$th term is the $v$-component $f_{v}\left(T^{-}\right)D_{v}\Phi\cdot Z\left(T^{+}\right)$,
the remaining $N-1$ components along the threshold reduce to $\Sigma_{i=1}^{N-1}f_{i}\left(T^{-}\right)Z_{i}\left(T^{-}\right)$
according to (\ref{eq:salt_bc_match}).
\begin{align*}
f_{v}\left(T^{-}\right)Z_{v}\left(T^{-}\right) & =\left(f_{v}\left(T^{-}\right)D_{v}\Phi-f_{v}\left(T^{-}\right)D_{v}\Phi-\Sigma_{i=1}^{N-1}\left(f_{i}\left(T^{-}\right)D_{i}\Phi\right)+f\left(T^{+}\right)\right)\cdot Z\left(T^{+}\right),\\
f_{v}\left(T^{-}\right)Z_{v}\left(T^{-}\right) & =-\Sigma_{i=1}^{N-1}f_{i}\left(T^{-}\right)Z_{i}\left(T^{-}\right)+f\left(T^{+}\right)\cdot Z\left(T^{+}\right),\\
f\left(T^{-}\right)\cdot Z\left(T^{-}\right) & =f\left(T^{+}\right)\cdot Z\left(T^{+}\right).
\end{align*}
Thus, we see that the final boundary condition does not depend on
the specific definition of the reset map $\Phi$ off the threshold
manifold, and simply enforces that the normalization condition $f\cdot Z=1$
holds across the reset.

\subsection{\label{subsec:PRC-phase}PRC phase shift for soft reset}

\begin{figure}[h]
\begin{centering}
\includegraphics[scale=0.8]{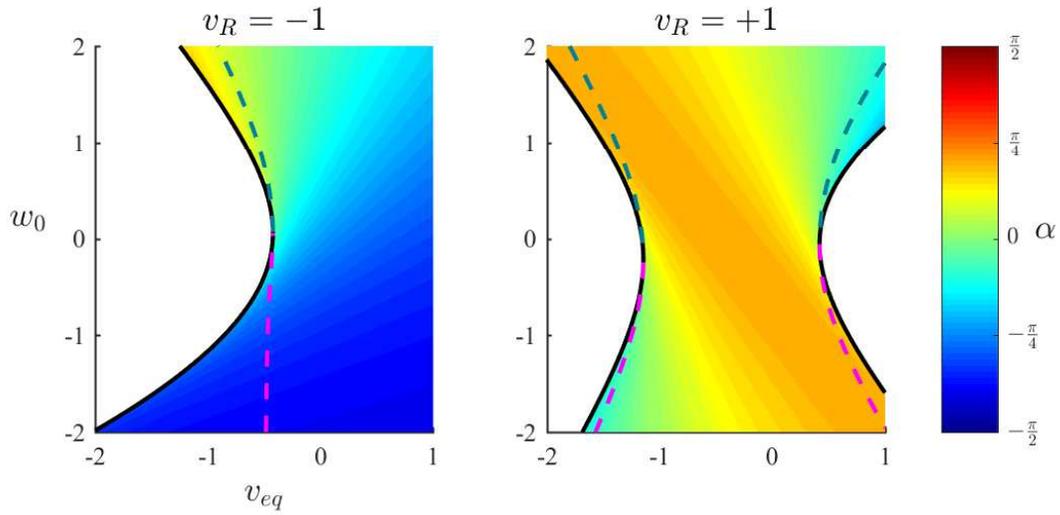}
\par\end{centering}
\caption{Phase shift $\alpha$ of the soft reset PRC $Z_{v}$, for $\lambda=0.1$.
(Note that $\alpha=0$ for hard reset.) Magenta and blue-green dashed
lines indicate stability boundaries of limit cycle for positive and
negative slope instabilities.\label{fig:Phase-shift}}
\end{figure}

The general form of the resonate-and-fire PRC is 
\[
Z_{v}\left(t\right)=\frac{A}{r_{0}}e^{\lambda t}\cos\left(t-T+\alpha\right),\ Z_{w}\left(t\right)=\frac{A}{r_{0}}e^{\lambda t}\sin\left(t-T+\alpha\right).
\]
The soft reset boundary condition determines the phase shift $\alpha$
as follows:

\begin{eqnarray*}
Z_{w}\left(T^{-}\right) & = & Z_{w}\left(0^{+}\right)\\
e^{\lambda T}\sin\left(\alpha\right) & = & \sin\left(\alpha-T\right)\\
\eta\sin T\cos\alpha & = & \left(\cos T-e^{\lambda T}\right)\sin\alpha\\
\alpha & = & \arctan\left(\frac{\sin T}{\cos T-e^{\lambda T}}\right).
\end{eqnarray*}
We show the phase shift evaluated over the full resonate-and-fire
parameter space in \figref{Phase-shift}.

\subsection{\label{subsec:app-Subthreshold-interaction-functio}Subthreshold
interaction function ($H_{sub}$) }

\begin{eqnarray*}
H_{sub}\left(\phi\right) & = & \frac{1}{T}\int_{0}^{T-\phi}Z_{v}\left(t\right)v_{sub}\left(t+\phi\right)dt+\frac{1}{T}\int_{T-\phi}^{T}Z_{v}\left(t\right)v_{sub}\left(t+\phi-T\right)dt-C\\
 & = & \frac{A}{2T}e^{-\lambda\phi}\left[\int_{0}^{T-\phi}\cos\left(\omega\left(t-T\right)+\alpha\right)\cos\left(\omega\left(t+\phi\right)+\theta_{0}\right)dt+\right.\\
 &  & \left....e^{\lambda T}\int_{T-\phi}^{T}\cos\left(\omega\left(t-T\right)+\alpha\right)\cos\left(\omega\left(t+\phi-T\right)+\theta_{0}\right)dt\right]-C\\
 & = & \frac{A}{2T}e^{-\lambda\phi}\left[e^{-\lambda T}\int_{0}^{T-\phi}\left(\cos\left(\omega\phi+\theta_{T}-\alpha\right)+\cos\left(\omega\left(2t+\phi-T\right)+\theta_{0}+\alpha\right)\right)dt+\right.\\
 &  & \left....\int_{T-\phi}^{T}\left(\cos\left(\omega\phi+\theta_{0}-\alpha\right)+\cos\left(\omega\left(2t+\phi-2T\right)+\theta_{0}+\alpha\right)\right)dt\right]-C\\
H_{sub}(\phi) & = & \frac{A}{2T}e^{-\lambda\phi}\left[\left(T-\phi\right)\cos\left(\omega\phi+\theta_{T}-\alpha\right)+\nicefrac{1}{\omega}\cos\left(\theta_{0}+\alpha\right)\sin\left(\omega\left(T-\phi\right)\right)+\right.\\
 &  & \left....e^{\lambda T}\phi\cos\left(\omega\phi+\theta_{0}-\alpha\right)+\nicefrac{1}{\omega}e^{\lambda T}\cos\left(\theta_{0}+\alpha\right)\sin\left(\omega\phi\right)\right]-C\\
C & = & \frac{1}{T}\int_{0}^{T}Z_{v}\left(t\right)v_{sub}\left(t\right)dt=\frac{A}{2T}\left(T\cos\left(\theta_{T}-\alpha\right)+\nicefrac{1}{\omega}\cos\left(\theta_{0}+\alpha\right)\sin\left(\omega T\right)\right)
\end{eqnarray*}

\subsection{\label{subsec:Slope-of-interaction}Slope of interaction function
components}

Here we evaluate the slope of each component of the resonate-and-fire
interaction function, its contribution to the slope of the odd component
of the interaction function, and expand the final result to first
order in the decay parameter $\lambda$.

\begin{eqnarray*}
C1^{\prime}(\phi) & = & \frac{1}{T}e^{-\lambda\phi}\left[-e^{\lambda T}\cos\left(T-\phi\right)-e^{\lambda T}\phi\sin\left(T-\phi\right)+\lambda e^{\lambda T}\phi\cos\left(T-\phi\right)...\right.\\
 &  & \left.+\cos\phi+\left(T-\phi\right)\sin\phi+\lambda\left(T-\phi\right)\cos\phi\right]\\
C1^{\prime}(0) & = & \frac{1}{T}\left(-e^{\lambda T}\cos T+1+\lambda T\right)\\
C1^{\prime}(T) & = & \frac{1}{T}\left(-1+\lambda T+e^{-\lambda T}\cos T\right)\\
C1_{odd}^{\prime}(0) & = & \lambda-\frac{1}{T}\cos T\sinh\left(\lambda T\right)\approx\lambda\left(1-\cos T\right)
\end{eqnarray*}
\begin{eqnarray*}
C2^{\prime}(\phi) & = & \frac{1}{T}e^{-\lambda\phi}\left[e^{\lambda T}\cos\phi-\lambda e^{\lambda T}\sin\phi-\cos\left(T-\phi\right)-\lambda\sin\left(T-\phi\right)\right]\\
C2^{\prime}(0) & = & \frac{1}{T}\left(e^{\lambda T}-\cos T-\lambda\sin T\right)\\
C2^{\prime}(T) & = & \frac{1}{T}\left(\cos T-\lambda\sin T-e^{-\lambda T}\right)\\
C2_{odd}^{\prime}(0) & = & \frac{1}{T}\left(\sinh(\lambda T)-\lambda\sin T\right)\approx\lambda\left(1-\frac{\sin T}{T}\right)
\end{eqnarray*}

\begin{eqnarray*}
S^{\prime}(\phi) & = & \frac{1}{T}e^{-\lambda\phi}\left[-e^{\lambda T}\sin\left(T-\phi\right)+e^{\lambda T}\phi\cos\left(T-\phi\right)-\lambda e^{\lambda T}\phi\sin\left(T-\phi\right)...\right.\\
 &  & \left.-\sin\phi+\left(T-\phi\right)\cos\phi-\lambda\left(T-\phi\right)\sin\phi\right]\\
S^{\prime}(0) & = & \frac{1}{T}\left(-e^{\lambda T}\sin T+T\right)\\
S^{\prime}(T) & = & \frac{1}{T}\left(T-e^{-\lambda T}\sin T\right)\\
S_{odd}^{\prime}(0) & = & 1-\frac{\sin T}{T}\cosh\left(\lambda T\right)\approx1-\frac{\sin T}{T}\left(1-\frac{\left(\lambda T\right)^{2}}{2}\right)
\end{eqnarray*}

\begin{eqnarray*}
H_{spike}^{\prime}(\phi) & = & Me^{\lambda(T-\phi)}\left[-\lambda\cos\left(\phi-\alpha\right)-\sin\left(\phi-\alpha\right)\right]\\
H_{spike}^{\prime}(0) & = & -Me^{\lambda T}\left[\lambda\cos\alpha-\sin\alpha\right]\\
H_{spike}^{\prime}(T) & = & -M\left[\lambda\cos\left(T-\alpha\right)+\sin\left(T-\alpha\right)\right]\\
H_{spike-odd}^{\prime}(0) & \approx & \frac{M}{2}\left[\sin\alpha+\sin\left(T-\alpha\right)+\lambda\left(T\sin\alpha-\cos\alpha+\cos\left(T-\alpha\right)\right)\right]
\end{eqnarray*}

\subsection{Amplitude of $H_{odd}$}

\begin{figure}[h]
\begin{centering}
\includegraphics[scale=0.8]{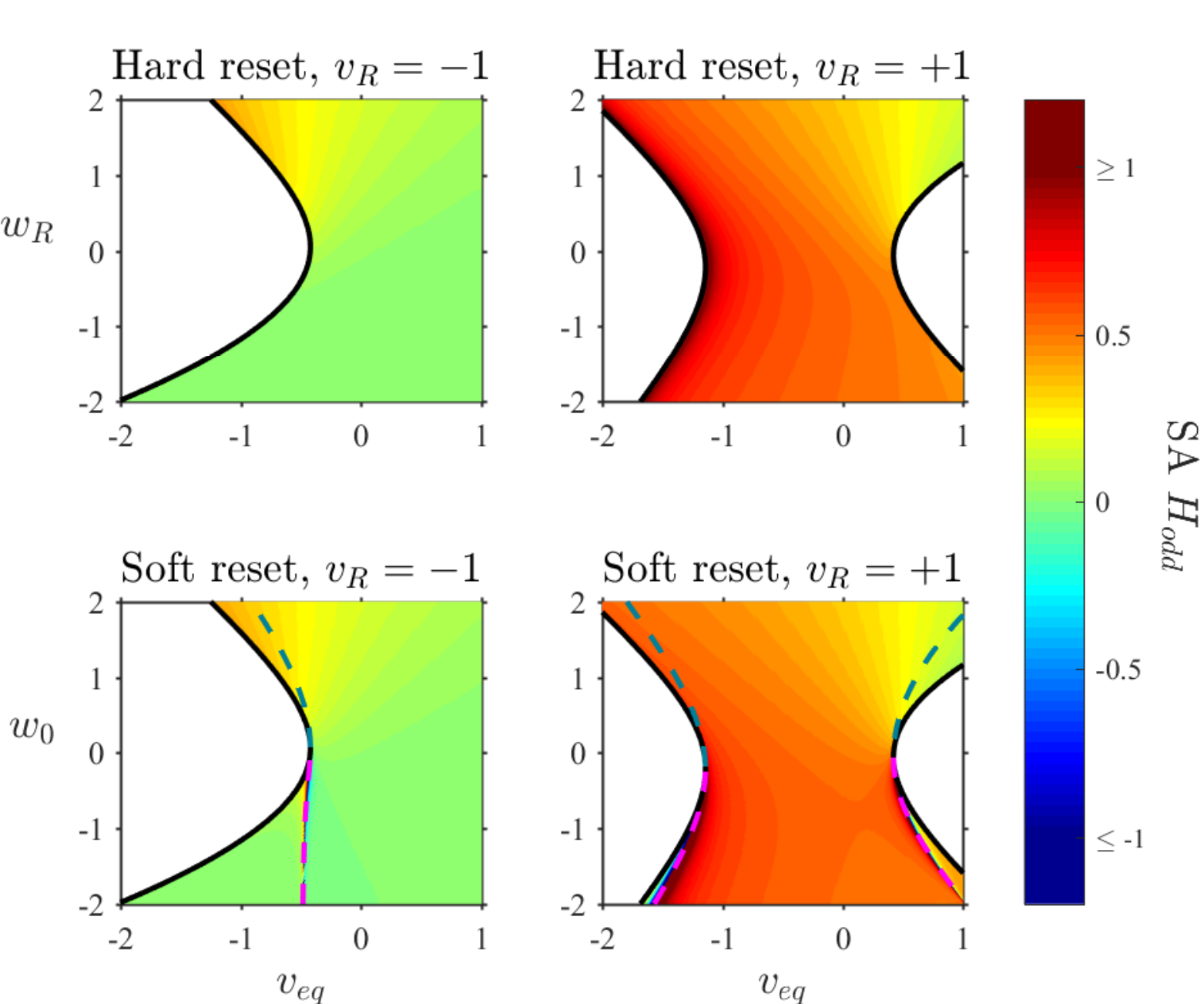}
\par\end{centering}
\caption{\label{fig:Signed-amp}Signed amplitude of $H_{odd}$, the odd component
of the subthreshold interaction function, for $\lambda=0.1$. Magenta
and cyan lines indicate stability boundaries of the limit cycle for
positive and negative slope instabilities.}
\end{figure}

Here we evaluate the \emph{signed amplitude,} 
\begin{align*}
\mathrm{SA}\left(H_{odd}\right) & =\mathrm{sign}\left(H_{odd}^{\prime}\left(0\right)\right)\max\left|H_{odd}\right|=H_{odd}\left(\phi_{max}\right),\\
 & \mathrm{where}\ \phi_{max}=\arg\max_{0\leq\phi\leq\nicefrac{T}{2}}\left|H_{odd}\left(\phi\right)\right|.
\end{align*}
Just as with the slope $H_{odd}^{\prime}\left(0\right)$, a larger
positive signed amplitude implies more robust near-synchronous phase
locking. We plot the signed amplitude of the resonate-and-fire interaction
function in \figref{Signed-amp}; for comparison, see the slope of
the interaction function in \figref{Aodd}. The slope and amplitude
are approximately equal, $\mathrm{SA}\left(H_{odd}\right)\approx\hat{H}_{odd}^{\prime}\left(0\right)=\frac{T}{2\pi}H_{odd}^{\prime}\left(0\right)$,
as expected from the Fourier approximation $\hat{H}_{odd}\left(\hat{\phi}\right)\propto\sin\left(\hat{\phi}\right)$.

\subsection{\label{subsec:App-Spike-interaction-function}Spike interaction function
effect $\Delta H_{spike}$}

\begin{figure}[H]
\begin{centering}
\includegraphics[scale=0.8]{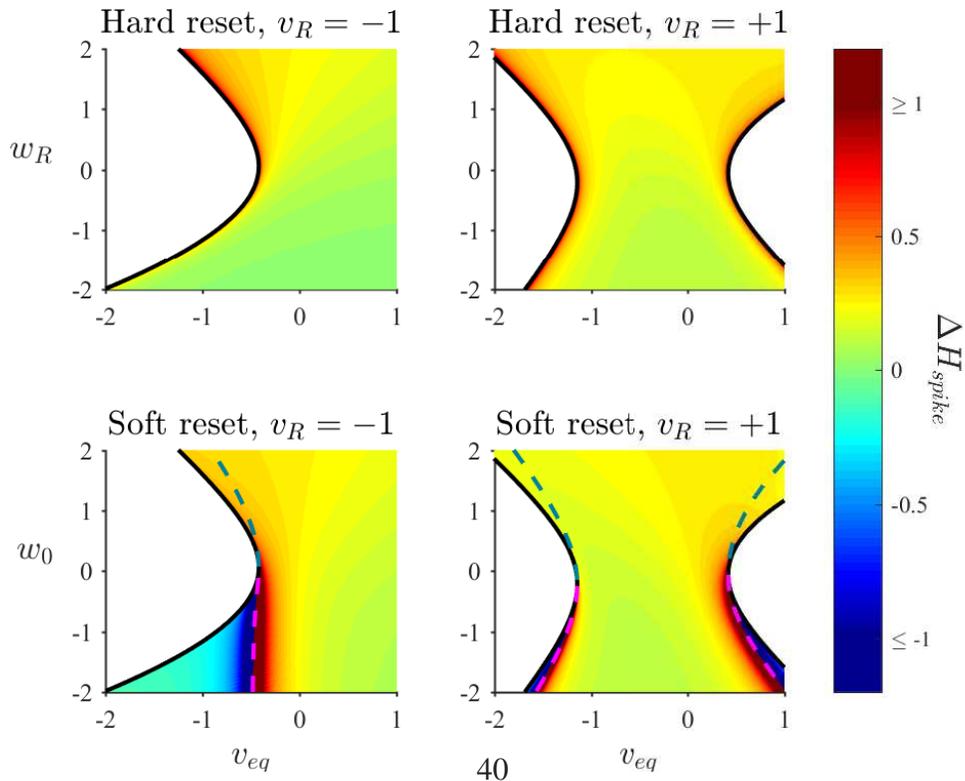}
\par\end{centering}
\caption{\label{fig:Discontinuity}Discontinuity of $H_{spike}$, the spike
component of the interaction function. Magenta and cyan lines indicate
stability boundaries of the limit cycle for positive and negative
slope instabilities. Parameters $\lambda=0.1$, $M=0.2$.}
\end{figure}

Here we assess the discontinuity of the spike interaction function,
\[
\Delta H_{spike}=H_{spike}\left(0^{+}\right)-H_{spike}\left(0^{-}\right)=\frac{M}{T}\left(Z(T^{-})-Z(0^{+})\right).
\]
We show the phase shift evaluated over the full resonate-and-fire
parameter space in \figref{Discontinuity}. We see that the discontinuity
is positive and relatively constant over the full parameter space,
increasing significantly only along the boundaries of the stable spiking
regime.

\section*{Acknowledgments}

This work was funded by the NDSEG fellowship (TC); NIH grants U01
HL126273 and SPARC A18-0491 (TL); and UC Davis Ophthalmology Research
to Prevent Blindness grant, a Simons Collaboration on the Global Brain
grant, and NIH grant R01 EY021581 (MG).

\bibliographystyle{siamplain2}
\bibliography{hybrid_resfire}

\end{document}